\documentclass[twocolumn,prl,showpacs,amsfonts,amsmath,eufrak,superscriptaddress]{revtex4-1}
\usepackage{amsmath,amsfonts,amssymb,mathtools,upgreek}
\usepackage{graphicx}
\usepackage[T1]{fontenc}

\usepackage{hyperref}\hypersetup{hidelinks}

\def\eq#1\en{\begin{equation}#1\end{equation}} 
\def\eqa#1\ena{\begin{align}#1\end{align}}
\def\eqg#1\eng{\begin{gather}#1\end{gather}}
\newcommand{\nl}{\notag\\}

\newcommand{\para}[1]{\medskip\par{\em #1}\/.---}
\newcommand{\prop}[1]{{\em #1}\/.---}

\newcounter{ct}

\newcommand{\ctl}[1]{\refstepcounter{ct}\thect\label{#1}}
\newcommand{\sbkt}[1]{\langle#1\rangle}
\newcommand{\bbkt}[1]{\bigl\langle#1\bigr\rangle}

\newcommand{\hell}{\hat{\ell}}
\newcommand{\ellL}{\ell_\mathrm{L}}
\newcommand{\ellR}{\ell_\mathrm{R}}
\newcommand{\hellL}{\hell_\mathrm{L}}
\newcommand{\hellR}{\hell_\mathrm{R}}

\newcommand{\LR}{^\mathrm{LR}}
\newcommand{\tv}{\tilde{v}}
\newcommand{\tmu}{\tilde{\mu}}
\newcommand{\bmu}{\bar{\mu}}
\newcommand{\Var}{\operatorname{Var}}
\newcommand{\calC}{\mathcal{C}}

\newcommand{\qedm}{\rule{1.5mm}{3mm}}
\newcommand{\dK}{d_\mathrm{K}}

\newcommand{\ourtitle}{Hierarchical Lorentz Mirror Model: Normal Transport and a Universal $2/3$ Mean--Variance Law}

\begin{document}
\title{\ourtitle}

\author{Rapha\"el Lefevere}\email[]{lefevere@lpsm.paris}
\affiliation{Laboratoire de Probabilit\'es, Statistiques et Mod\'elisation, Universit\'e Paris Cit\'e}

\author{Hal Tasaki}\email[]{hal.tasaki@gakushuin.ac.jp}
\affiliation{Department of Physics, Gakushuin University, Mejiro, Toshima-ku, Tokyo 171-8588, Japan}

\date{\today}

\begin{abstract}

The Lorentz mirror model provides a clean setting to study macroscopic transport generated solely by quenched environmental randomness.
We introduce a hierarchical version whose distribution of left--right crossings satisfies an exact recursion.
In dimensions $d\ge3$, we prove two-sided bounds that support normal transport: the mean conductance scales as (cross-section)/(length).
A Gaussian closure, supported by numerics, predicts that the variance-to-mean ratio of the dimensionless conductance converges to the universal value $2/3$ for all $d\ge2$ (the ``$2/3$ law'').
We provide numerical evidence for the $2/3$ law in the original (non-hierarchical) Lorentz mirror model in $d=3$, and conjecture that it is a universal signature of normal transport induced by random current matching.
In the marginal case $d=2$, our hierarchical recursion reproduces the known scaling of the mean and variance of conductance.
\par\noindent
{\footnotesize
A YouTube video discussing the background and the main results of the paper is available: 
\par\noindent\url{https://youtu.be/G1nqKd6MiXo}}
\end{abstract}

\maketitle

Deriving macroscopic normal transport---as described by diffusive laws such as Fick's, Ohm's, and Fourier's laws---from microscopic deterministic dynamics remains a central challenge in nonequilibrium statistical mechanics \cite{BonettoLebowitzReyBellet,LepriLiviPoliti,DharReview,LebowitzSpohn}.
Lorentz-type models, in which particles move deterministically through an externally imposed (often disordered) environment, provide a natural testing ground for this program \cite{Lorentz1905,DettmannReview,vanBeijeren,Basile,DharDhar1999}.
The Lorentz mirror model introduced by Ruijgrok and Cohen \cite{RuijgrokCohen} is a particularly clean lattice realization: local scattering is deterministic, so any normal transport must be generated by quenched environmental randomness rather than by stochastic forcing or chaotic microscopic dynamics \cite{RuijgrokCohen,ZiffKongCohen1991,Basile,KraemerSanders,BunimovichTroubetzkoy,BunimovichTroubetzkoy2,NienhuisRietman,CohenWang,OwczarekPrellberg,BezuidenhoutGrimmett1999,Lefevere2015,ChiffaudelLefevere,ElboimGloriaHernandez,Lefevere2025,Grimmett,KozmaSidoravicius,Ryan2021,GrimmettKesten,NahumSernaSomozaOrtuno2013}.
See, e.g., \cite{NahumSernaSomozaOrtuno2013} and references therein for its close relation to loop models.

In the present Letter, we propose and study a hierarchical version of the Lorentz mirror model.
We rigorously establish normal transport in dimensions $d\ge3$;
in the marginal case $d=2$, the Gaussian closure recovers the weakly
anomalous transport with a logarithmic correction found in
\cite{NahumSernaSomozaOrtuno2013}.
Most strikingly, we conjecture that the ratio of the sample-to-sample variance of the dimensionless conductance to its mean approaches the universal value $2/3$ not only in the hierarchical model but also in the original mirror model, suggesting a novel universal signature of normal transport induced by random current matching.
A  video discussing the main results of the paper is available \cite{video}.

\begin{figure}
\centerline{\includegraphics[width=.8\linewidth]{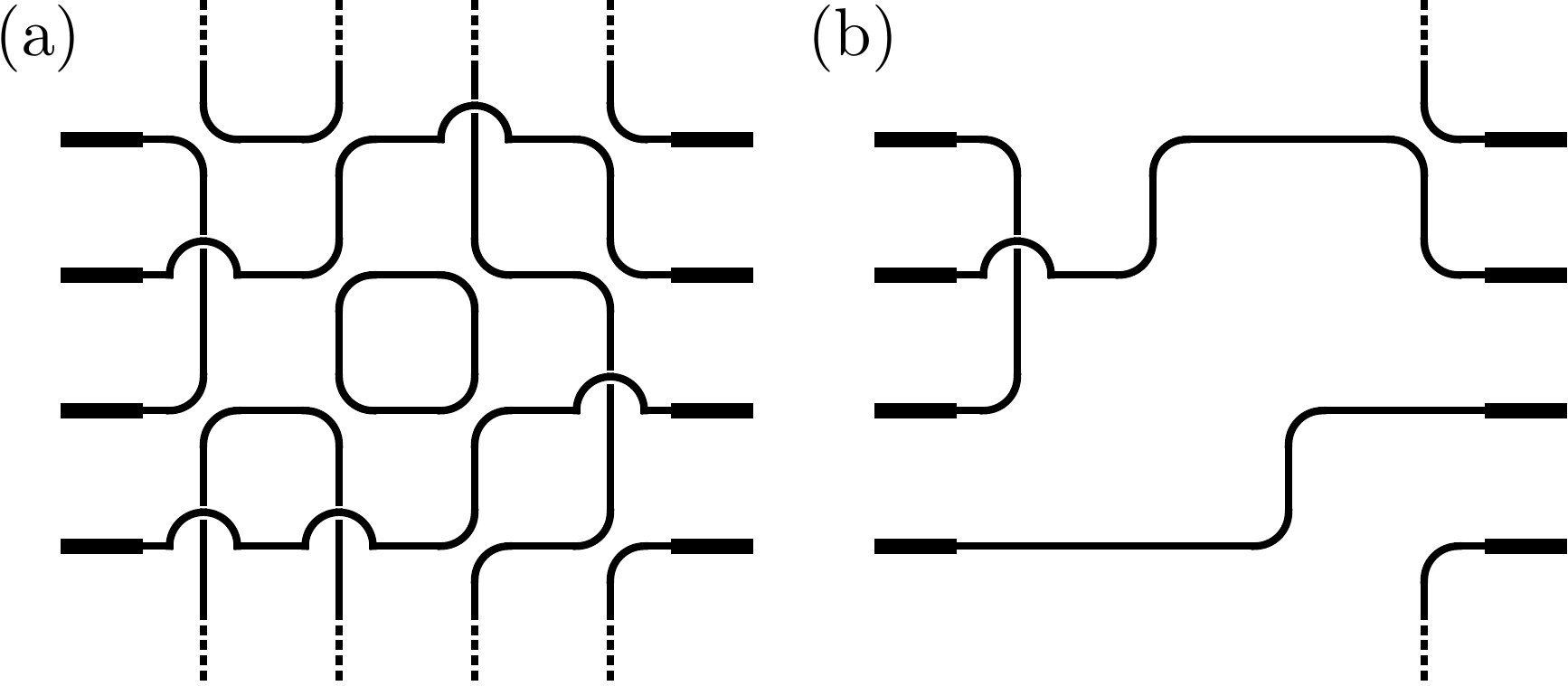}}
\caption{
Original Lorentz mirror model in $d=2$ (periodic boundary conditions in the vertical direction).
(a)~A random choice of local pairings yields a collection of disjoint trajectories.
(b)~The induced perfect matching of the external edges (thick lines).
There are two crossings.
}
\label{f:org}
\end{figure}

\para{Original Lorentz mirror model}
Let us describe the Lorentz mirror model, starting with $d=2$.
The model may be viewed as a special case of the compactly packed loop model with crossings \cite{NahumSernaSomozaOrtuno2013}.
Consider the $L\times L$ square lattice with periodic boundary conditions in the vertical direction and open boundary conditions in the horizontal direction.
We attach one external edge to each boundary vertex on the left and on the right (thick lines in Fig.~\ref{f:org}), so that every vertex has four incident edges.
At each vertex, we independently choose one of the three pairings of the four incident edges, each with probability $1/3$.
These local pairings decompose the edges into a collection of disjoint trajectories and induce a global perfect matching of the external edges; see Fig.~\ref{f:org}.
As the name suggests, in $d=2$ the local pairing may be viewed as arising from reflections by randomly placed mirrors \cite{SM}.

Similarly, for $d\ge3$, we consider the $d$-dimensional $L\times\cdots\times L$ hypercubic lattice with open boundary conditions in one (horizontal) direction and periodic boundary conditions in the remaining $d-1$ directions, and attach $L^{d-1}$ external edges to each of the two open boundaries.
(This particular geometry, suited to the study of transport properties, was studied in \cite{Lefevere2015}.)
We then independently choose one of the $(2d-1)!!$ pairings of the $2d$ incident edges with equal probability at each vertex.
(Note that this does not literally correspond to mirror configurations.)
Again, the local wiring induces a global perfect matching of the external edges.

Given such a matching of the external edges, a matched pair is said to be a crossing if it connects the left boundary to the right boundary.
We denote by $\calC$ the number of crossings in a given random environment.
We interpret $\calC$ as dimensionless conductance, which is related to dimensionful (measurable) conductance by $G=\gamma\,\calC$ with a constant $\gamma$ \cite{SM}.
The expectation value $\bmu(L)=\sbkt{\calC}$ is called the mean conductance. 
This terminology is motivated by the dynamical interpretation: a particle (or light ray) arriving at a vertex along an incident edge leaves along its paired edge, so trajectories are uniquely determined by the environment.
In a nonequilibrium setting where particle baths maintain different densities at the two open boundaries, the stationary current is proportional to $\calC$ \cite{Lefevere2015,ChiffaudelLefevere}.
See SM.1 of \cite{SM} for details.

Normal transport implies that the mean conductance is proportional to the cross-section $A=L^{d-1}$ of the system divided by its horizontal length $L$, i.e.,
\eq
\bmu(L)\propto A/L=L^{d-2}.
\label{eq:Fick}
\en
For $d=3$, the scaling \eqref{eq:Fick} is supported by numerical results in \cite{ChiffaudelLefevere} and is consistent with the multiscale analysis of \cite{Lefevere2025} based on a closure assumption on trajectory correlations.

Note that \eqref{eq:Fick} would follow readily if the particle motion were diffusive in the usual sense (e.g., well approximated by a random walk).
However, the microscopic motion in the Lorentz mirror model is far from diffusive: for large $L$ a substantial fraction of vertices lie on closed orbits \cite{BunimovichTroubetzkoy,ElboimGloriaHernandez}; see also Fig.~\ref{f:org}~(a).
Nevertheless, any particle injected through an external edge necessarily exits through an external edge by construction.
A full probabilistic understanding of the model---and a rigorous derivation of \eqref{eq:Fick}---appears to be a formidable task.
We nonetheless expect that, on large scales, the model exhibits universal features that are captured by the idealized coarse-grained model introduced in this Letter.

\para{The hierarchical model}
We follow the tradition of hierarchical models in statistical mechanics \cite{Dyson,BleherSinai,GallavottiKnops,ColletEckmann,BerkerOstlund,GawedzkiKupiainen1980,KaufmanGriffiths,GawedzkiKupiainen1981,GawedzkiKupiainen1982}
and introduce a hierarchical version of the Lorentz mirror model, in which coarse-graining is implemented explicitly and is amenable to analysis.
Our construction is inspired by the multiscale ``slab concatenation'' viewpoint developed for the original Lorentz mirror model in \cite{Lefevere2025}.
See also \cite{DharRajeshStilck2011} for a similar construction.

We define the model in dimensions $d=1,2,\ldots$.
Fix a positive even integer $A_0$.
A generation-$n$ block ($n=0,1,2,\ldots$) has
$A_n:=A_0\,2^{(d-1)n}$
external edges on each of its left and right sides, and horizontal length
$L_n:=2^n$.

\begin{figure}
\centerline{\includegraphics[width=.8\linewidth]{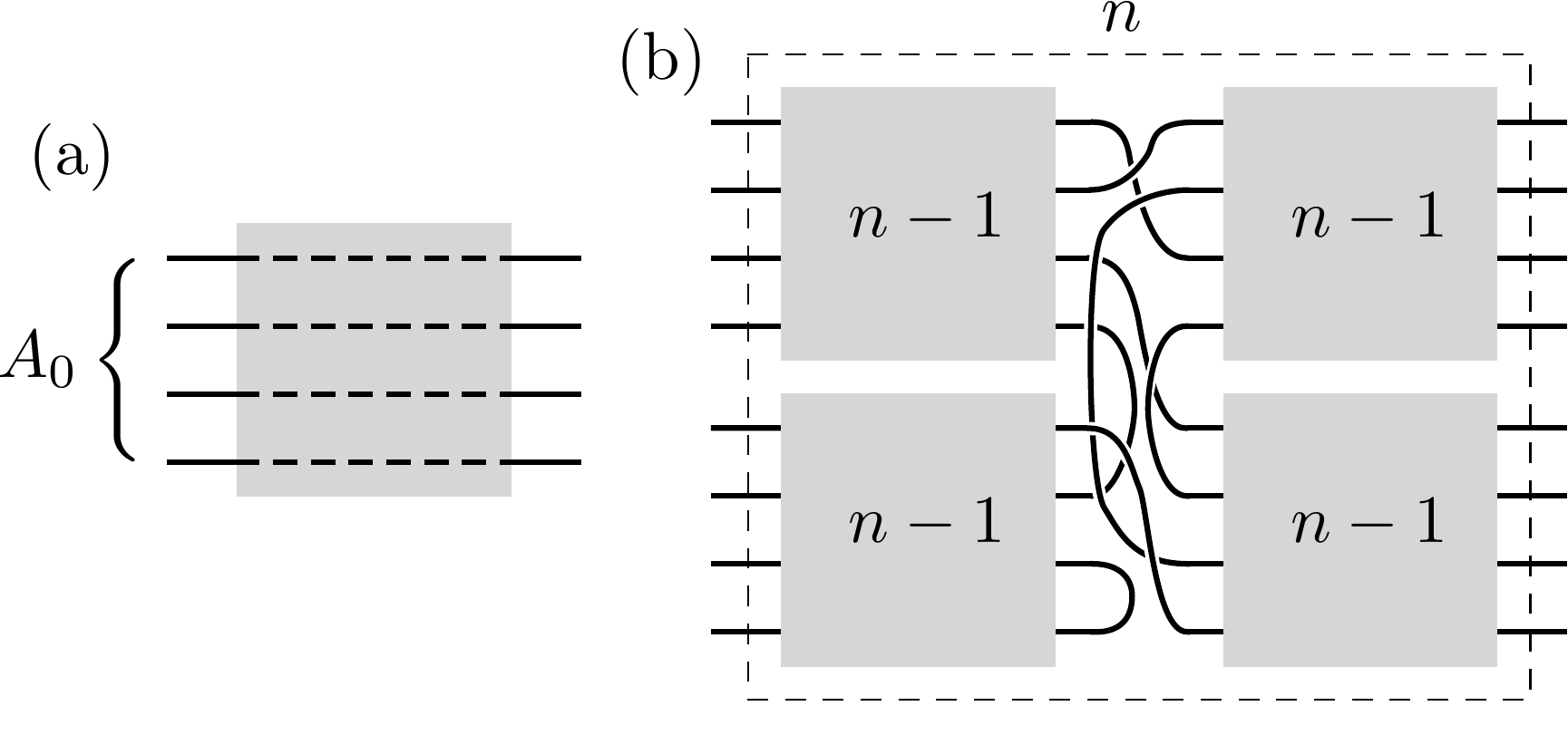}}
\caption{
(a)~The generation-$0$ block.
(b)~The construction of the generation-$n$ block from $2^d$ independent copies of the generation-$(n-1)$ block.
}
\label{f:hierarchical}
\end{figure}

The generation-$0$ block is defined by matching each left external edge with a right external edge, so that it has exactly $A_0$ crossings; see Fig.~\ref{f:hierarchical}~(a).
For $n\ge 1$, the generation-$n$ block is constructed from $2^d$ independent copies of the generation-$(n-1)$ block, arranged as a $2\times\cdots\times2$ array in $d$ dimensions, as illustrated in Fig.~\ref{f:hierarchical}~(b).
We group $2^{d-1}$ copies in parallel to form the left half and $2^{d-1}$ copies in parallel to form the right half.
The $A_n=2^{d-1}A_{n-1}$ external edges on the far left (respectively, far right) become the left (respectively, right) external edges of the generation-$n$ block.
The remaining $2A_n=2^{d}A_{n-1}$ external edges at the interface between the two halves are then paired uniformly at random, i.e., we choose one of the $(2A_n-1)!!$ perfect matchings with equal probability.
This completes the generation-$n$ block; see Fig. \ref{f:hierarchical} (b).

The internal random wiring induces a perfect matching of the $2A_n$ external edges of the generation-$n$ block.
Again, a crossing is a matched pair connecting opposite sides.
Note that the number of crossings is always even, since the matching at the interface preserves the parity.

Let $P_n(\ell)$, where $\ell\in2\mathbb{Z}_{\ge0}$, denote the probability that the generation-$n$ block has exactly $\ell$ crossings.
For any $f(\ell)$, we write $\sbkt{f(\hell)}_n:=\sum_\ell f(\ell)P_n(\ell)$, where $\hell$ denotes the (random) number of crossings in the generation-$n$ block.
In particular, we set $\mu_n:=\sbkt{\hell}_n$ and call it the mean conductance.

The distribution $P_n$ is determined by the initial condition $P_0(\ell)=\delta_{\ell,A_0}$ and, for $n\ge1$, the recursion
\eq
P_n(\ell)=\sum_{\ell_1,\ldots,\ell_{2^d}} K(\ell\,|\,\ellL,\ellR)\,\prod_{j=1}^{2^d}P_{n-1}(\ell_j),
\label{eq:rec}
\en
where the sum is over nonnegative even $\ell_1,\ldots,\ell_{2^d}$, and
$\ellL=\sum_{j=1}^{2^{d-1}}\ell_j$ and $\ellR=\sum_{j=2^{d-1}+1}^{2^d}\ell_j$ are the numbers of crossings in the left and right halves, respectively.
The kernel $K(\ell\,|\,\ellL,\ellR)$ is the conditional probability that the random matching at the interface produces exactly $\ell$ left--right pairs among the $\ellL+\ellR$ interface edges that are endpoints of crossings in the generation-$(n\!-\!1)$ subblocks, and is given by
\eq
K(\ell\,|\,\ellL,\ellR)=\frac{(\ellL-\ell-1)!!(\ellR-\ell-1)!!}{(\ellL+\ellR-1)!!}\binom{\ellL}{\ell}\binom{\ellR}{\ell}\ell!,
\label{eq:kernel}
\en
if $\ell\in S(\ellL,\ellR)$, and $K(\ell\,|\,\ellL,\ellR)=0$ otherwise, where
$S(a,b):=2\mathbb{Z}\cap[0,\min\{a,b\}]$.
It is normalized as $\sum_{\ell\in S(\ellL,\ellR)}K(\ell\,|\,\ellL,\ellR)=1$.
This follows by observing that the uniform random matching of all $2A_n$ interface edges induces, by permutation symmetry, a uniform random matching of the $\ellL+\ellR$ distinguished endpoints.

For fixed $\ellL$ and $\ellR$, the conditional mean and variance of $\hell$ are
\eq
\tmu(\ellL,\ellR):=\sum_{\ell\in S(\ellL,\ellR)}
\ell\,K(\ell\,|\,\ellL,\ellR)=\frac{\ellL \ellR}{\ellL+\ellR-1},
\label{eq:condmean}
\en
and
\eq
\tv(\ellL,\ellR)
=\frac{2\ellL \ellR(\ellL-1)(\ellR-1)}{(\ellL+\ellR-3)(\ellL+\ellR-1)^2}.
\en
See End Matter and \cite{SM}.

\para{Detailed derivation of \eqref{eq:kernel}}
To derive \eqref{eq:kernel}, fix wirings inside the generation-$(n\!-\!1)$ subblocks that have $\ellL$ and $\ellR$ crossings in the left and right halves.
Among the $A_n$ interface edges adjacent to the left half, exactly $\ellL$ are endpoints of crossings in the left half, while the remaining $A_n-\ellL$ interface edges are paired within the left half (``U-turns'').
Similarly, among the $A_n$ interface edges adjacent to the right half, $\ellR$ are endpoints of crossings and $A_n-\ellR$ belong to U-turns.
The uniform random perfect matching of the $2A_n$ interface edges, together with these fixed U-turn matchings inside the two halves, induces a matching of the $\ellL+\ellR$ distinguished endpoints (those belonging to crossings): starting from such an endpoint, follow the interface matching, then (if necessary) follow the prescribed U-turn in that half, and continue until another distinguished endpoint is reached.
By permutation symmetry, the induced matching is uniform among all perfect matchings of the $\ellL+\ellR$ distinguished edges.

There are $(\ellL+\ellR-1)!!$ perfect matchings of $\ellL+\ellR$ labeled endpoints.
To obtain exactly $\ell$ left--right pairs among them, we choose the $\ell$ endpoints on each side that participate in left--right pairs ($\binom{\ellL}{\ell}\binom{\ellR}{\ell}$ choices), pair them across ($\ell!$ choices), and pair the remaining $\ellL-\ell$ left endpoints and $\ellR-\ell$ right endpoints among themselves to form U-turns (in $(\ellL-\ell-1)!!$ and $(\ellR-\ell-1)!!$ ways, respectively, with the convention $(-1)!!=1$).
Dividing by $(\ellL+\ellR-1)!!$ yields \eqref{eq:kernel}.

\para{Normal transport}
When $\ellL,\ellR\gg1$, \eqref{eq:condmean} reduces to a simple suggestive form 
$1/\tmu(\ellL,\ellR)\simeq(1/\ellL)+(1/\ellR)$,
which is reminiscent of the series-resistance formula.
Since the dominant contribution to the recursion \eqref{eq:rec} should come from
$\ellL\simeq\ellR\simeq2^{d-1}\mu_{n-1}$,
this suggests the recursion $\mu_n\simeq2^{d-2}\mu_{n-1}$ for the mean conductance.
Its solution gives
$\mu_n\simeq(2^{d-2})^n\mu_0=A_n/L_n$,
which is precisely the normal-transport scaling \eqref{eq:Fick}.

Our main mathematical contribution is a rigorous version of this observation in $d\ge3$.

\prop{Theorem}
Let $d\ge3$.
For sufficiently large $A_0$, there are positive constants $C_d$ and $C'_d$ such that
\eq
C_d\frac{A_n}{L_n}-1\le\mu_n\le C'_d\frac{A_n}{L_n},
\label{eq:mainbounds}
\en
for all $n$.
If  $A_0\gg1$, one has $C_d\simeq C'_d\simeq1$, and hence $\mu_n\simeq A_n/L_n$ for all $n$.

See End Matter for precise constants and the proof.
For $d=3$, any strictly positive even integer $A_0$ is allowed, and we have $C_3=1-5/(4A_0)$ and $C'_3=1+1/(3A_0)$.

As we shall discuss below, we expect logarithmic growth of $\mu_n$ in the marginal dimension $d=2$.
We only have an upper bound consistent with the conjecture.  See Theorem~\ref{upper} in End Matter.
For $d=1$, the same $A_0/L_n$ scaling can be proved up to the scale where $\mu_n$ becomes $O(1)$; see Proposition~\ref{lower1d} in End Matter.

\para{Gaussian heuristic and the $2/3$ law}
A useful (and surprisingly accurate) heuristic is obtained by a Gaussian closure \cite{SM}:
(i)~assume that, for sufficiently large $n$, the probability distribution $P_n$ is well approximated by a Gaussian with mean
$\mu_n=\sbkt{\hell}_n$ and variance $v_n:=\sbkt{(\hell-\mu_n)^2}_n$;
(ii)~approximate the interface kernel $K(\ell\,|\,\ellL,\ellR)$ by a Gaussian in $\ell$,
\eq
K(\ell\,|\,\ellL,\ellR)\propto
\exp\!\Bigl[-\frac{\{\ell-\tmu(\ellL,\ellR)\}^2}{2\,\tv(\ellL,\ellR)}\Bigr],
\label{eq:Kgauss}
\en
with the conditional mean $\tmu(\ellL,\ellR)$ and conditional variance $\tv(\ellL,\ellR)$ given above.
Our exact numerics \cite{SM} confirm (i) via the rapid decay of the Kolmogorov distance between the standardized $P_n$ and the standard normal distribution, while (ii) can be proved in an appropriate asymptotic regime.

Under this closure, the recursion \eqref{eq:rec} essentially reduces to Gaussian integrals, and for $d\ge3$ yields the leading recursions 
\eq
\mu_n\simeq 2^{d-2}\mu_{n-1},\quad
v_n\simeq 2^{d-3}\mu_{n-1}+2^{d-4}v_{n-1}.
\label{eq:gaussrec}
\en
Here, the two contributions to $v_n$ come from the conditional variance in \eqref{eq:Kgauss} and from the fluctuations of $\ellL$ and $\ellR$, respectively.

Remarkably, the ratio $v_n/\mu_n$ satisfies the recursion
\eq
\frac{v_n}{\mu_n}\simeq \frac{1}{2}+\frac{1}{4}\frac{v_{n-1}}{\mu_{n-1}},
\label{eq:ratio23}
\en
which is independent of $d$.
Therefore, $v_n/\mu_n$ rapidly converges to $2/3$ as $n\to\infty$ (the ``$2/3$ law''), in agreement with our exact numerical iteration of the recursion \cite{SM}.


\para{Gaussian closure in $d=2$}
In the marginal dimension $d=2$, the leading drift in \eqref{eq:gaussrec} cancels; retaining the next-order terms predicts a logarithmic growth
\eq
\mu_n\simeq \frac{n}{12}=\frac{\log L_n}{12\log2},
\label{eq:2D}
\en
i.e., a weakly anomalous behavior.  Note that the transport is extremely weak.
The Gaussian closure still yields the 2/3 law, $v_n/\mu_n\to2/3$.
These conclusions, as well as the (slow) approach of $P_n$ to a Gaussian shape, are supported by numerics \cite{SM}.

Remarkably, both the logarithmic behavior of the mean conductance and the $2/3$ law were obtained for the original Lorentz mirror model in $d=2$ by Nahum, Serna, Somoza, and Ortu\~no~\cite{NahumSernaSomozaOrtuno2013} via a 
nonlinear sigma-model analysis and large-scale Monte Carlo simulations (up to $L=10^6$).
See also \cite{ZiffKongCohen1991,OwczarekPrellberg} for earlier numerical indications of logarithmic corrections.

\begin{figure}[t]
\centering
\includegraphics[width=\linewidth]{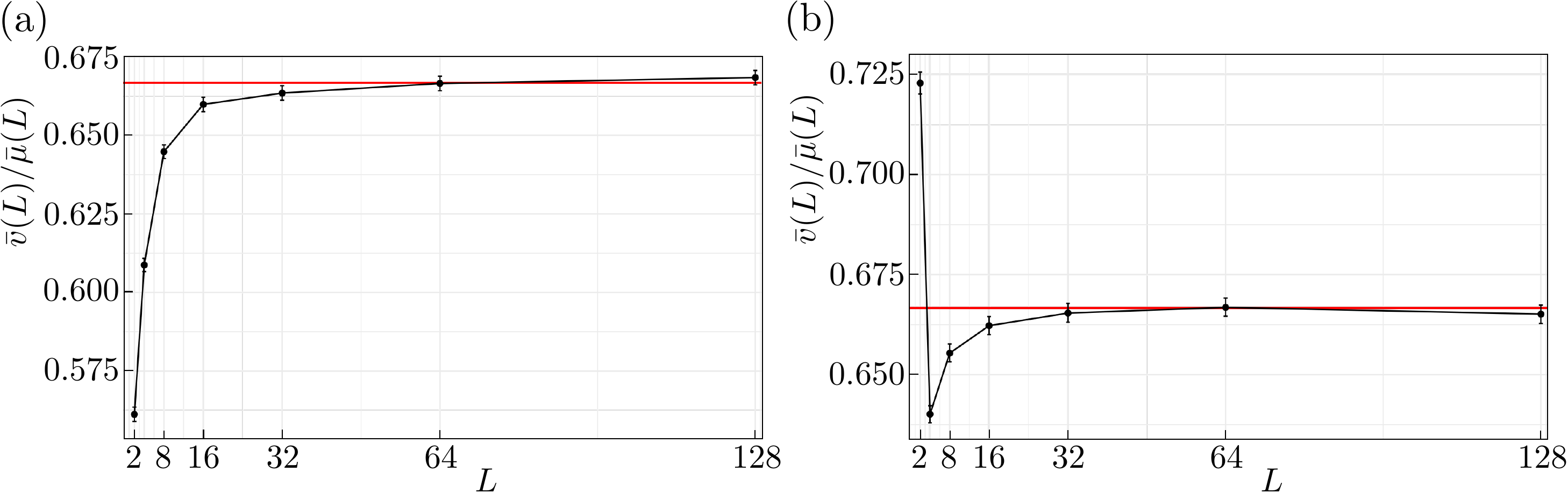}
\caption{
Ratio $\bar{v}(L)/\bmu(L)$ of the sample-to-sample variance $\bar{v}(L)$ to the mean $\bmu(L)$ of the number of crossings, as a function of $L=2^n$ with $n=1,\ldots,7$.
Error bars indicate $95\%$ confidence intervals.
(a)~Standard rule (all local pairings allowed).
(b)~Orthogonal rule (restricted local pairings; no edge is paired with its opposite).
The horizontal line marks $2/3$.
}
\label{fig:ratio3D_mirrors_panel}
\end{figure}

\para{Numerical results for the original Lorentz mirror model in $d=3$}
The rapid stabilization of $v_n/\mu_n$ to $2/3$ in the hierarchical model for $d\ge2$, together with the RG analysis and large-scale simulations in \cite{NahumSernaSomozaOrtuno2013} for the original model in $d=2$, motivates viewing the variance-to-mean ratio of the 
dimensionless conductance $\calC$ as a universal amplitude ratio, characterizing a universality class that contains the present hierarchical Lorentz mirror model.
(See SM.1 of \cite{SM} for the implication for a dimensionful conductance.)
We shall now see that this expectation is justified, in a stronger sense than we initially anticipated.

Normal-transport scaling \eqref{eq:Fick} in the original Lorentz mirror model for $d=3$ was first supported numerically in \cite{ChiffaudelLefevere}.
The proportionality constant (conductivity) $\kappa:=\lim_{L\uparrow\infty}\bmu(L)/L$ was derived in \cite{Lefevere2025} from a multiscale analysis under a closure hypothesis on trajectory correlations.
This constant is nonuniversal and depends on the microscopic pairing rule.

Here, we report preliminary simulations of the original (non-hierarchical) model aimed at testing the variance-to-mean ratio of the conductance. 
The results are already strongly suggestive: the ratio
$\bar v(L)/\bar\mu(L)$ rapidly approaches the value $2/3$ as $L$
increases and appears insensitive to microscopic details of the local
pairing rule; see Fig.~\ref{fig:ratio3D_mirrors_panel}.
Simulations at a different aspect ratio also support the $2/3$ law
\cite{SM}.
This unexpected robustness leads us to conjecture that the ``$2/3$ law'' persists well beyond the hierarchical setting.
It is natural to interpret $\bar{v}(L)/\bmu(L)$ as a universal amplitude ratio for transport problems in which coarse-graining effectively randomizes the matching of conserved currents.

In our simulations, we considered the Lorentz mirror model on the $L\times L\times L$ cubic lattice with two different local pairing rules.
The first is the standard rule discussed at the beginning of the Letter.
The second is an ``orthogonal'' rule, in which at each vertex we choose uniformly among the eight local pairings such that no edge is paired with its opposite (equivalently, the velocity of a particle always turns by $90^\circ$ at each visited vertex).
As we recall at the end of End Matter, the $d=2$ model with the orthogonal rule belongs to a different universality class.

For each $L=2^n$ with $n=1,\ldots,7$ and for each rule, we generated $6.4\times 10^5$ independent quenched random environments (configurations of local pairings).
For a given environment, we faithfully evaluated the conductance $\calC$, i.e., the number of left--right crossings, by tracing the deterministic trajectories of particles injected from all $L^2$ external edges on the left boundary and recording their exit edges.
Repeating this procedure over all the environments yields estimates of the mean $\bmu(L)=\sbkt{\calC}$, the variance $\bar{v}(L):=\sbkt{(\calC-\bmu(L))^2}$, and the ratio $\bar{v}(L)/\bmu(L)$ with $95\%$ confidence intervals.
See \cite{SM}.

\para{Discussion}
We introduced a hierarchical Lorentz mirror model that captures deterministic transport generated solely by quenched environmental randomness through random matching.
For this model, we proved that the mean conductance exhibits normal-transport scaling, even though the microscopic dynamics is far from diffusive and many trajectories are closed, as in the original mirror model.
More precisely, we established that $\mu_n\propto A_n/L_n$  in all dimensions $d\ge3$.
The marginal case $d=2$ remains open: the Gaussian approximation and numerics suggest that $\mu_n$ grows logarithmically with the system size, in agreement with the previous study of the non-hierarchical model \cite{NahumSernaSomozaOrtuno2013}.

A second outcome is the ``$2/3$ law'': whenever $\mu_n$ grows with scale, the variance satisfies $v_n/\mu_n\to2/3$.
In the hierarchical model, this emerges from the Gaussian closure leading to the recursion \eqref{eq:ratio23}, and is supported numerically.
The same ratio 2/3 was already found in the non-hierarchical model with $d=2$ in \cite{NahumSernaSomozaOrtuno2013}.
Moreover, we provide numerical evidence that the same ratio holds for the original Lorentz mirror model in $d=3$, and appears robust under variations of the local pairing rule and under a change of the aspect ratio.
This motivates the conjecture that the ``$2/3$ law'' is universal within a broad class of random-current-matching models, but it need not extend to generic Ohmic or thermal conductance; see Ref.~\cite{Pal2026} for a contrasting random-resistor example.

Our results suggest that $d=2$ is the unique marginal dimension of the hierarchical model, and we conjecture the same for the original model.
A single injected trajectory may be viewed as a non-backtracking walk with strong memory (encoding the quenched random environment), suggesting a comparison with ``true'' (myopic) self-avoiding walks and related self-repelling processes, for which $d=2$ is also marginal and which become diffusive in $d\ge3$ \cite{AmitParisiPeliti,Toth1995,TothWerner1998,HorvathTothVeto2012}.
At the same time, if crossings were produced by essentially independent diffusive trajectories, one would expect Poisson counting statistics, for which $v/\mu$ equals $1$ rather than approaching the sub-Poisson value $2/3$.
This contrast suggests that the global current matching---rather than the scaling of an individual trajectory---is the mechanism behind normal transport with the $2/3$ law.
A rigorous derivation of the $2/3$ law, and control of the marginal case $d=2$, remain important open problems.

Our hierarchical Lorentz mirror model provides a tractable yet nontrivial
setting for deterministic transport in a quenched random environment.
From a broader physics viewpoint, the ``$2/3$ law'' may serve as a
diagnostic of an underlying universality class.
A natural extension of the field-theoretic analysis of \cite{NahumSernaSomozaOrtuno2013} suggests that this class is
described by the Goldstone phase of an appropriate nonlinear sigma model,
with the crossing statistics controlled by its renormalized stiffness.
It is particularly intriguing to determine whether the same long-distance
description applies to the non-hierarchical Lorentz mirror model and emerges in
continuous-space models such as Lorentz's original model
\cite{Lorentz1905}.

\medskip
\begin{acknowledgments}
We thank Hosho Katsura for pointing out that Ref.~\cite{NahumSernaSomozaOrtuno2013} contains results closely related to the present study.
We also thank the anonymous referees for their constructive comments, which helped us improve the manuscript.
R.L. thanks Gakushuin University for its hospitality during his stay in Tokyo, when the main part of this work was carried out.
He also acknowledges the financial support from FJ-LMI IRL 2025 CNRS-The University of Tokyo, which made the visit possible.
The present research is supported by JSPS Grants-in-Aid for Scientific Research Nos. 22K03474 (R.L. and H.T.) and 25K07171 (H.T.).

We made extensive use of ChatGPT (versions~5.1 and~5.2, Plus/Pro) throughout this project for code development, assistance with proofs and calculations, reference searches, and drafting of the manuscript.
In particular, the role of the identity \eqref{eq:ELiab} in the proof of Theorem~\ref{lowergen} was identified with the assistance of ChatGPT~5.2~Pro (although, interestingly, it never realized \eqref{eq:ELiab} until we pointed out that the quantity could be evaluated).
We also used ChatGPT 5.6 Sol Pro in revising the manuscript in response to the referee reports.
\end{acknowledgments}


\newpage
\appendix

\section*{End Matter}
We give a complete proof of our main theorem on normal transport, splitting it into Theorems~\ref{upper} and \ref{lowergen}.

\para{Conditional expectation values}
We begin by proving two exact identities, \eqref{eq:ELab} and \eqref{eq:ELiab}, for conditional expectations.
Recall that the kernel is normalized as $\sum_{\ell\in S(a,b)}K(\ell\,|\,a,b)=1$ for any nonnegative even $a$ and $b$.
The combinatorial interpretation of \eqref{eq:kernel} also shows that $\sum_\ell K(\ell\,|\,a,b)=1$ when both $a$ and $b$ are odd and positive, where the sum is over odd $\ell\in[1,\min\{a,b\}]$.
It is also useful to rewrite \eqref{eq:kernel} as
\eqa
K(\ell\,|\,a,b)=\frac{a!\,b!}{(a+b-1)!!\,(a-\ell)!!\,(b-\ell)!!\,\ell!}.
\label{eq:kernel2}
\ena

Let $a$ and $b$ be even and positive.
Using the above normalization in the odd case (applied to $a-1$ and $b-1$), we have
\eqa
1&=\sum_{\ell\in S(a,b)\backslash\{0\}}K(\ell-1\,|\,a-1,b-1)
\nl
&=\sum_{\ell\in S(a,b)\backslash\{0\}}\frac{(a-1)!\,(b-1)!}{(a+b-3)!!\,(a-\ell)!!\,(b-\ell)!!\,(\ell-1)!}
\nl
&=\frac{a+b-1}{ab}\sum_{\ell\in S(a,b)}\ell\,K(\ell\,|\,a,b).
\ena
This yields the equality in
\eq
\sum_{\ell\in S(a,b)}
\ell\,K(\ell\,|\,a,b)=\frac{ab}{a+b-1}
\le\frac{a+b}{4}+\frac{1}{3},
\label{eq:ELab}
\en
which proves \eqref{eq:condmean}.
To prove the upper bound in \eqref{eq:ELab}, assume $ab\neq0$ and set $S:=a+b\ge4$.
Then
\eq
\frac{ab}{a+b-1}-\frac{a+b}{4}
=\frac{S-(a-b)^2}{4(S-1)}
\le\frac{S}{4(S-1)}\le\frac{1}{3}.
\en
If $a=0$ or $b=0$, the left-hand side of \eqref{eq:ELab} vanishes and the bound is trivial.

Similarly, for even and positive $a$ and $b$, we find
\eqa
1&=\sum_{\ell\in S(a,b)}K(\ell+1\,|\,a+1,b+1)
\nl
&=\sum_{\ell\in S(a,b)}\frac{(a+1)!\,(b+1)!}{(a+b+1)!!\,(a-\ell)!!\,(b-\ell)!!\,(\ell+1)!}
\nl
&=\frac{(a+1)(b+1)}{a+b+1}\sum_{\ell\in S(a,b)}\frac{1}{\ell+1}\,K(\ell\,|\,a,b),
\ena
which leads to
\eqa
&\sum_{\ell\in S(a,b)}
\frac{1}{\ell+1}\,K(\ell\,|\,a,b)
=\frac{a+b+1}{(a+1)(b+1)}
\nl&=(a+1)^{-1}+(b+1)^{-1}-\{(a+1)(b+1)\}^{-1}.
\label{eq:ELiab}
\ena

\para{Upper bounds}
For any function $f(a,b)$, define
\eq
\bbkt{f(\hellL,\hellR)}\LR_n
:=\sum_{\ell_1,\ldots,\ell_{2^d}} f(\ellL,\ellR)\,\prod_{j=1}^{2^d}P_{n}(\ell_j),
\label{eq:exp_n}
\en
where the sum is over nonnegative even $\ell_1,\ldots,\ell_{2^d}$, and
$\ellL=\sum_{j=1}^{2^{d-1}}\ell_j$ and $\ellR=\sum_{j=2^{d-1}+1}^{2^d}\ell_j$.
With this notation, the recursion \eqref{eq:rec} reads
$P_n(\ell)=\sbkt{K(\ell\,|\,\hellL,\hellR)}\LR_{n-1}$.
Recalling $\mu_n=\sum_{\ell}\ell\,P_n(\ell)$ and using the upper bound in \eqref{eq:ELab}, we obtain
\eq
\mu_n
\le\frac{\bbkt{\hellL+\hellR}\LR_{n-1}}{4}+\frac{1}{3}
=2^{d-2}\,\mu_{n-1}+\frac{1}{3},
\en
for $n=1,2,\ldots$, where we used
$\sbkt{\hellL}\LR_{n-1}=\sbkt{\hellR}\LR_{n-1}=2^{d-1}\mu_{n-1}$.
For $d\ne2$, we rewrite this as
$(\mu_n+C''_d)\le2^{d-2}(\mu_{n-1}+C''_d)$ with
$C''_d=\{3(2^{d-2}-1)\}^{-1}$, and obtain the following (remember that $\mu_0=A_0$).

\prop{Theorem \ctl{upper}}
For any $n=0,1,\ldots$, we have
\eq
\mu_n\le
\begin{cases}
2^{(d-2)n}(\mu_0+C''_d)-C''_d
=C'_d\,A_n/L_n-C''_d, & d\ne2;\\
A_0+\dfrac{n}{3}
=A_0+(3\log 2)^{-1}\log L_n, & d=2,
\end{cases}
\label{eq:munupper}
\en
with constants $C'_d=1+C''_d/A_0$.

For $d\ge3$, where $C''_d>0$, this implies the upper bound in \eqref{eq:mainbounds}.
Note that the bound for $d=2$ is consistent with the expected logarithmic growth \eqref{eq:2D}.
The bound for $d\ne2$ is also valid for $d=1$ with $C''_1=-2/3$.

\para{Lower bounds}
To prove lower bounds, we introduce
\eq
\eta_n:=\sbkt{(\hell+1)^{-1}}_n=\sum_\ell(\ell+1)^{-1}P_n(\ell).
\label{eq:etadef}
\en
Since $x\mapsto x^{-1}$ is convex, Jensen's inequality yields
$\eta_n=\sbkt{(\hell+1)^{-1}}_n\ge1/\sbkt{\hell+1}_n$, and hence
$\mu_n\ge\eta_n^{-1}-1$.
Thus an upper bound on $\eta_n$ implies a lower bound on $\mu_n$.

Moreover, \eqref{eq:ELiab} implies the exact identity
\eqa
\eta_n
=&\bbkt{(\hellL+1)^{-1}}\LR_{n-1}+\bbkt{(\hellR+1)^{-1}}\LR_{n-1}
\nl&-\bbkt{(\hellL+1)^{-1}}\LR_{n-1}\bbkt{(\hellR+1)^{-1}}\LR_{n-1},
\label{eq:etan_exact}
\ena
where we noted that $\hellL$ and $\hellR$ are independent in the expectation \eqref{eq:exp_n}.

As a warm-up, we start from the less interesting case with $d=1$.
Dropping the last (negative) term in \eqref{eq:etan_exact} and noting
$\sbkt{(\hellL+1)^{-1}}\LR_{n-1}=\sbkt{(\hellR+1)^{-1}}\LR_{n-1}=\eta_{n-1}$,
we obtain $\eta_n\le2\eta_{n-1}$, hence
$\eta_n\le2^n\eta_0=2^n/(A_0+1)$.
We have thus proved the following.

\prop{Proposition \ctl{lower1d}}
For $d=1$ and any $n=0,1,\ldots$,
\eq
\mu_n\ge(A_0+1)/L_n-1.
\label{eq:munlower1d}
\en

Combined with the upper bound \eqref{eq:munupper}, this shows
$\mu_n\simeq A_0/L_n$ for $d=1$, as long as $A_0\gg1$ and $\mu_n$ remains much larger than unity.
This may be regarded as (a weak form of) normal transport in the limited length scale.

For $d\ge3$, we prove a lower bound with the correct scaling in the same spirit.
Let us first discuss a rough argument and state the theorem.
Define
\eq
u_n:=\bbkt{(\hellL+1)^{-1}}\LR_{n}=\bbkt{(\hellR+1)^{-1}}\LR_{n},
\label{eq:un}
\en
where the equality follows from symmetry.
(Note that $u_n=\eta_n$ when $d=1$.)
Then \eqref{eq:etan_exact} implies the identity
\eq
\eta_n=2u_{n-1}-(u_{n-1})^2.
\label{eq:eta=u}
\en
A further estimate based on concavity (proved below) yields
\eq
u_n\le\frac{\eta_n}{2^{d-1}-(2^{d-1}-1)\eta_n}.
\en
Combining this with \eqref{eq:eta=u}, we arrive at the recursive inequality
\eq
\eta_n\le2^{2-d}\,\eta_{n-1}+(1-2^{1-d})^2(\eta_{n-1})^2.
\en
Iterating this inequality and using $\mu_n\ge\eta_n^{-1}-1$, we obtain the following.

\prop{Theorem \ctl{lowergen}}
For $d\ge3$, one has
\eq
\mu_n\ge C_d\,A_n/L_n-1,
\label{eq:munlowergend}
\en
for any $n=0,1,\ldots$, with a constant
\eq
C_d=1-\frac{1}{A_0}\biggl(\frac{(2^{d-1}-1)^2}{2(2^{d-1}-2)}-1\biggr),
\en
where we take $A_0$ sufficiently large so that $C_d>0$.

\para{Proof of Theorem~\ref{lowergen}}
We discuss the case $d=3$; the extension to general $d\ge3$ is straightforward \cite{SM}.
From \eqref{eq:un} and \eqref{eq:exp_n}, we have
\eq
u_n=\sum_{\ell_1,\ldots,\ell_4}\Bigl(1+\sum_{j=1}^{4}\ell_j\Bigr)^{-1}\prod_{j=1}^{4}P_n(\ell_j),
\label{eq:un3}
\en
where the sum is over nonnegative even $\ell_1,\ldots,\ell_4$ (we have summed over $\ell_5,\ldots,\ell_8$).
Make the change of variables $y=(\ell+1)^{-1}\in(0,1]$ and define $Q_n(y)=P_n(y^{-1}-1)$.
Then $\eta_n=\sum_y y\,Q_n(y)$ and \eqref{eq:un3} becomes
\eq
u_n=\sum_{y_1,\ldots,y_4}\Bigl(\sum_{j=1}^4y_j^{-1}-3\Bigr)^{-1}\prod_{j=1}^4Q_n(y_j).
\en
Fix $y_2,y_3,y_4\in(0,1]$ and set $c:=\sum_{j=2}^4y_j^{-1}-3\ge0$.
The function $(y_1^{-1}+c)^{-1}=y_1/(1+c y_1)$ is concave in $y_1\in(0,1]$, since its second derivative equals $-2c(1+c y_1)^{-3}\le0$.
By Jensen's inequality,
\eq
u_n\le\sum_{y_2,y_3,y_4}\Bigl(\eta_n^{-1}+\sum_{j=2}^4y_j^{-1}-3\Bigr)^{-1}\prod_{j=2}^4Q_n(y_j),
\en
and repeating the same argument for $y_2,y_3,y_4$ yields
$u_n\le\bigl(4\eta_n^{-1}-3\bigr)^{-1}=\eta_n/(4-3\eta_n)$.
Substituting this (with $n\mapsto n-1$) into \eqref{eq:eta=u}, we find
\eq
\eta_n
\le\frac{8\eta_{n-1}-7\eta_{n-1}^2}{(4-3\eta_{n-1})^2}.
\en
Noting that
\eq
\frac{8\eta-7\eta^2}{(4-3\eta)^2}-\Bigl(\frac{\eta}{2}+\frac{9}{16}\eta^2\Bigr)
=-\frac{\eta^2(9\eta-8)^2}{16(4-3\eta)^2}\le0,
\en
we obtain the recursive inequality
\eq
\eta_n\le\frac{\eta_{n-1}}{2}+\frac{9}{16}\eta_{n-1}^2.
\label{eq:eta3drec}
\en

It remains to iterate \eqref{eq:eta3drec}.
Set $x_n=2^n\eta_n$.
Then \eqref{eq:eta3drec} implies
\eq
x_n\le x_{n-1}+\frac{9}{8}2^{-(n-1)}x_{n-1}^2,
\en
and therefore
\eq
\frac{1}{x_n}\ge\frac{1}{x_{n-1}}-\frac{9}{8}2^{-(n-1)},
\en
since $(x+ax^2)^{-1}\ge x^{-1}-a$ for $x>0$ and $a\ge0$.
Summing over $n$ and using $x_0=\eta_0=(A_0+1)^{-1}$, we obtain
\eq
\frac{1}{x_n}\ge\frac{1}{x_0}-\frac{9}{8}\sum_{k=0}^{n-1}2^{-k}
\ge (A_0+1)-\frac{9}{4}=A_0-\frac{5}{4}.
\en
Thus $\eta_n^{-1}\ge (A_0-\frac{5}{4})2^n$, and hence
$\mu_n\ge\eta_n^{-1}-1\ge (A_0-\frac{5}{4})L_n-1$, which is the desired bound \eqref{eq:munlowergend}.~\qedm

\para{The orthogonal rule in $d=2$}
We finally recall a known special case in $d=2$.
If the straight-through pairing is forbidden and, at each vertex, either
turning pairing is chosen with probability $1/2$, the trajectories give the
standard medial-lattice hull representation of critical bond percolation.
With Cardy's convention for spanning clusters, our dimensionless conductance
$\calC$ is related to the number $N_c$ of clusters connecting the two open
boundaries by $\calC=2N_c$.
\cite{ZiffKongCohen1991,NahumSernaSomozaOrtuno2013}.
For our square-cylinder geometry, Cardy's Coulomb-gas formula for the
scaling-limit annulus distribution \cite{CardyAnnulus2002} yields
$\Var(\calC)/\sbkt{\calC}\simeq1.30$, clearly distinct from the $2/3$ law.
Moreover, at the percolation point this ratio depends on the cylinder aspect
ratio, whereas the $2/3$ law conjectured here is expected to be independent
of the aspect ratio.


\clearpage
\onecolumngrid 

\setcounter{page}{1}
\renewcommand{\thepage}{SM-\arabic{page}}

\setcounter{equation}{0}
\renewcommand{\theequation}{SM.\arabic{equation}}
\newcounter{equationStore}

\setcounter{figure}{0}
\renewcommand{\thefigure}{SM.\arabic{figure}}
\setcounter{table}{0}
\renewcommand{\thetable}{SM.\arabic{table}}

\setcounter{section}{0}
\renewcommand{\thesection}{SM.\arabic{section}}
\renewcommand{\thesubsection}{SM.\arabic{section}.\arabic{subsection}}

\newcommand{\secnumfont}{\bfseries}
\renewcommand{\section}[1]{
\setcounter{equationStore}{\value{equation}}
\refstepcounter{section}\par\bigskip\noindent{\secnumfont\thesection\quad #1}\par\medskip
\setcounter{equation}{\value{equationStore}}
}
\renewcommand{\subsection}[1]{\par\medskip\refstepcounter{subsection}\par\smallskip\noindent{\itshape\thesubsection\quad #1}\par\smallskip}

\noindent{\large\bf Supplemental Material for ``\ourtitle''}

\begin{center}
Rapha\"el Lefevere and Hal Tasaki
\end{center}

\begin{quotation}
In this Supplemental Material, we collect background information, straightforward but somewhat involved computations and extensions, as well as details of our numerical simulations.
These materials are intended for interested readers; the main text of the Letter, together with the End Matter, is self-contained.

Specifically, we discuss the dimensionless and dimensionful conductances (Sec.~\ref{sm:conductance-units}), the relation between the traditional ``mirror'' formulation and the local-pairing formulation used in the main text (Sec.~\ref{sm:mirror}), details of the Gaussian closure (Sec.~\ref{sm:Gauss}), the extension of the proof of Theorem~\ref{lowergen} to general dimensions $d\ge3$ (Sec.~\ref{sm:Theorems3}), and additional information on our numerical simulations (Sec.~\ref{SM:NumericsHierarchical}).
\end{quotation}

\section{Dimensionless and dimensionful conductance}
\label{sm:conductance-units}

\begin{figure}[b]
\centering
\includegraphics[width=.35\textwidth]{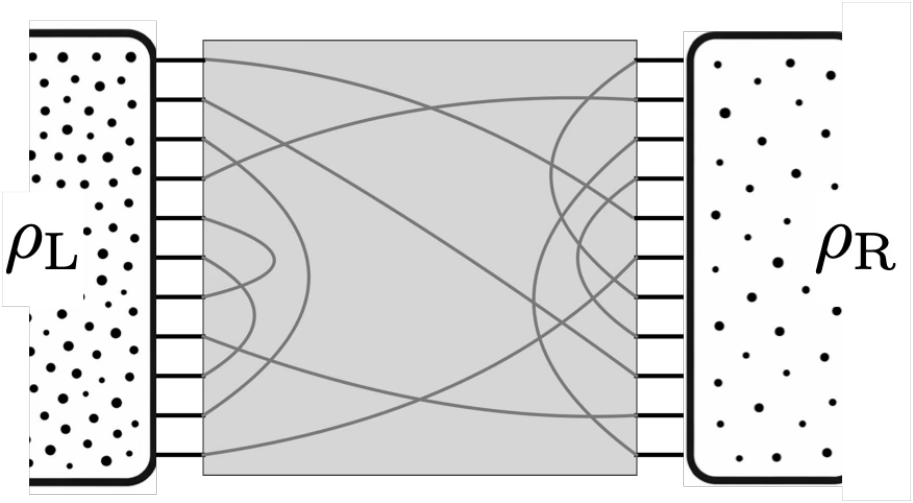}
\caption{
Schematic reservoir setup.  All external edges on side
$s=\mathrm{L},\mathrm{R}$ are coupled to the reservoir of density $\rho_s$.
The gray curves schematically represent the trajectories that induce the
matching of the external edges; those joining opposite sides are crossings.
For identical reservoir couplings, the dimensionful particle conductance is
$G=\gamma\,\calC$, where $\gamma$ is the conductance of one perfectly
transmitting channel.
}
\label{fig:Reservoirs}
\end{figure}

The crossing number $\calC$ introduced in the main text is an integer and is
therefore dimensionless; its mean and variance are dimensionless as well.  It
may be interpreted as the conductance measured in units of a single perfectly
transmitting channel.  We make this interpretation concrete by considering the
standard setup in which the external edges are coupled to particle reservoirs,
as illustrated in Fig.~\ref{fig:Reservoirs}.  See also
Refs.~\cite{Lefevere2015,ChiffaudelLefevere}.

Fix a realization of the random environment with $\calC$ left--right
crossings.  All external edges on side $s=\mathrm{L},\mathrm{R}$ are coupled
identically to the reservoir on that side.  Let $\rho_s$ denote the reservoir
density normalized by a fixed reference density, so that $\rho_s$ is
dimensionless, and suppose that particles are injected into each external edge on side
$s$ at a constant mean rate $r(\rho_s)$, with the same injection-rate
function $r$ on the two sides.
The particles are noninteracting.  Upon
injection, each particle is assigned a positive speed drawn from an arbitrary
distribution, and then moves
ballistically along the deterministic path specified by the local pairings
until it reaches another external edge and is absorbed by the corresponding
reservoir.  We assume that the mean traversal time is finite, so that the
stationary state has a finite mean number of particles in the system.

The speed distribution affects the traversal times and the stationary particle
number, but not the stationary mean particle-number current.  To see this,
consider a left--right crossing $a$ and let $N_{a,s}(t)$ denote the number of
particles on $a$ that were injected from side $s=\mathrm{L},\mathrm{R}$.  Let
$\mathbb{E}_{\mathrm{dyn}}$ denote the average over the injection process and
the speeds for this fixed environment.  If $j_{a,s}^{\mathrm{in}}(t)$ and
$j_{a,s}^{\mathrm{out}}(t)$ denote the corresponding mean injection and exit
rates, particle conservation gives
\eq
\frac{d}{dt}\,\mathbb{E}_{\mathrm{dyn}}[N_{a,s}(t)]
=
j_{a,s}^{\mathrm{in}}(t)-j_{a,s}^{\mathrm{out}}(t).
\label{sm:eq:path-continuity}
\en
Consequently, in the stationary state,
$j_{a,s}^{\mathrm{out}}=j_{a,s}^{\mathrm{in}}=r(\rho_s)$, independently of
the speed distribution.

Each left--right crossing therefore carries a left-to-right flux
$r(\rho_{\mathrm L})$ and a right-to-left flux $r(\rho_{\mathrm R})$.  A trajectory joining two external edges on the same side produces no net transfer
between the reservoirs.  Thus the stationary net particle current, taken
positive from left to right, is exactly
\eq
J=\{r(\rho_{\mathrm L})-r(\rho_{\mathrm R})\}\,\calC.
\label{sm:eq:current-general}
\en
For the simple linear coupling $r(\rho)=\gamma\rho$, where $\gamma>0$ has the
dimension of inverse time, the current can be written as
\eq
J=G(\rho_{\mathrm L}-\rho_{\mathrm R}),
\qquad
G=\gamma\,\calC.
\label{sm:eq:G-gammaC}
\en
Here $G$ is the dimensionful particle conductance, with
$[G]=[\gamma]=T^{-1}$.  The linear injection law is only a convenient
idealization.  More generally, for differentiable $r(\rho)$, write
$\rho_{\mathrm L,\mathrm R}=\rho_0\pm\Delta\rho/2$.  In linear response,
Eq.~\eqref{sm:eq:current-general} gives
$J=r'(\rho_0)\Delta\rho\,\calC+o(\Delta\rho)$, and hence
$G=r'(\rho_0)\calC$.  In this more general setting, $\gamma$ below is
understood as the linear-response single-channel conductance $r'(\rho_0)$.

Let $\sbkt{\cdot}$ and $\Var$ denote the expectation and variance over
the quenched random environment, with the reservoir coupling held fixed.
Equation~\eqref{sm:eq:G-gammaC} gives
\eq
\sbkt{G}=\gamma\sbkt{\calC},
\qquad
\Var(G)=\gamma^2\Var(\calC),
\label{sm:eq:G-moments}
\en
and hence the $2/3$ law can be written in terms of the dimensionful
conductance as
\eq
\frac{\Var(G)}{\gamma\sbkt{G}}
=
\frac{\Var(\calC)}{\sbkt{\calC}}
\longrightarrow\frac{2}{3}.
\label{sm:eq:physical-23}
\en
The universal statement is the dimensionless ratio in
Eq.~\eqref{sm:eq:physical-23}, normalized by the single-channel conductance
$\gamma$.  Equivalently, $\Var(G)/\sbkt{G}\to(2/3)\gamma$; this latter ratio
has the dimension of inverse time and is not itself universal.

It is crucial to note that the single-channel scale $\gamma$ depends solely on the coupling between the particle bath and the external edges.
It can in principle be calibrated using a
reference sample with the same reservoir coupling.  If all $A$ left external
edges are connected to right external edges, then $\calC_{\mathrm{ref}}=A$ and one obtains the constant from measured reference conductance as $\gamma=G_{\mathrm{ref}}/A$.
Thus the prediction can be expressed entirely in terms of conductance
measurements on an ensemble of test samples and on the reference sample.

Finally, we stress that $\Var(G)$ denotes the sample-to-sample variance of the stationary
conductance: one first determines $G$ from the stationary mean current for each
fixed environment and then takes the variance over environments.  It is
distinct from temporal particle-counting noise in a fixed environment.

\section{From mirrors to local pairings}
\label{sm:mirror}
Here, we explain the precise relation, in $d=2$, between the traditional
``mirror'' formulation of the Lorentz mirror model and the ``tube'' (local-pairing) formulation
used in the main text.  

\begin{figure}[b]
\centering
\includegraphics[width=.60\textwidth]{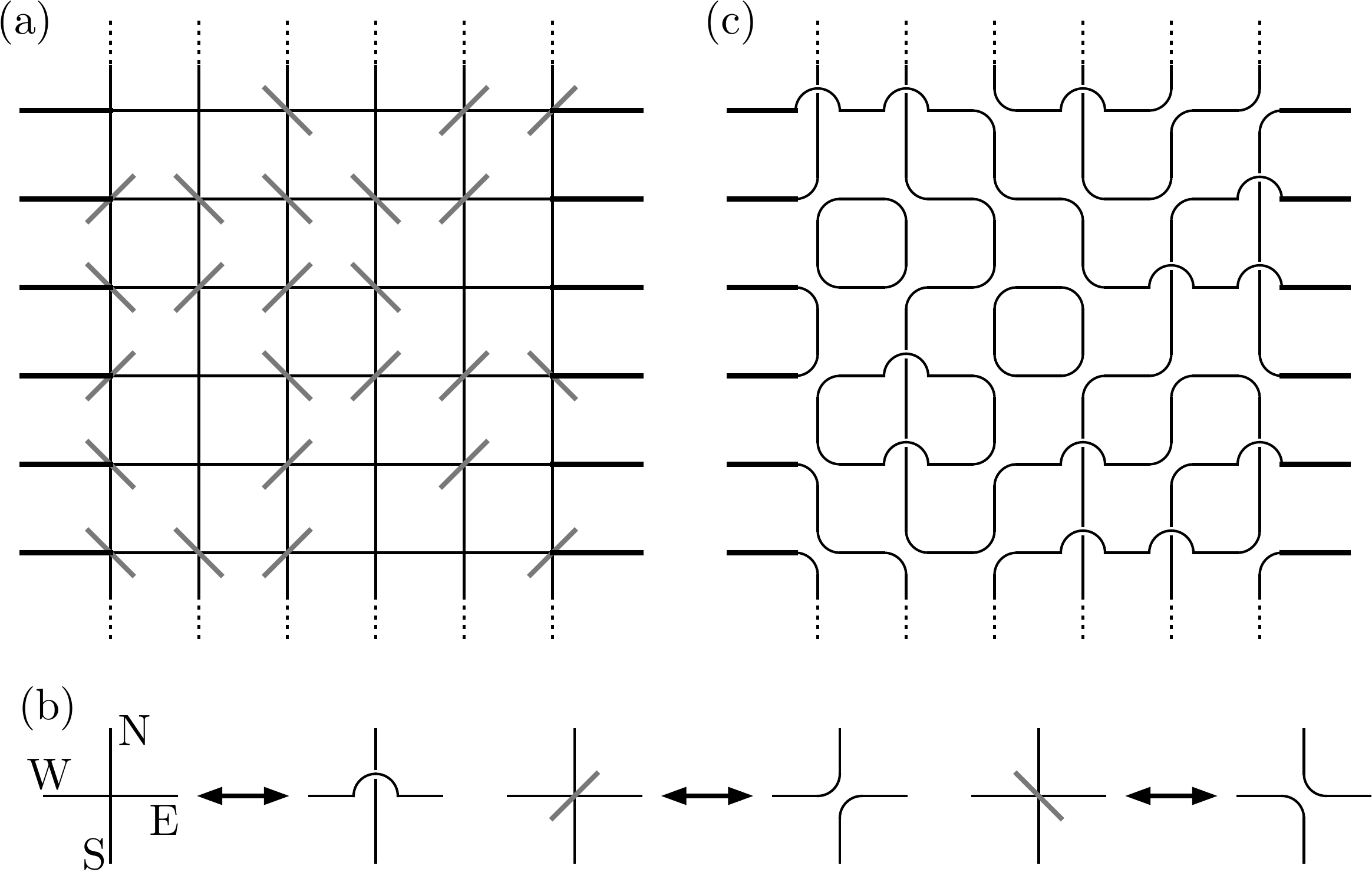}
\caption{
Equivalence between the mirror and tube representations in $d=2$.
(a)~A configuration of mirrors (empty sites, ``/'' mirrors, and ``\textbackslash'' mirrors) on the $L\times L$ square lattice with periodic boundary conditions vertically and external edges attached on the left and right.
(b)~Local correspondence between the three vertex types (empty, ``/'' mirror, and ``\textbackslash'' mirror) and the three pairings of the four incident edges.
(c)~The corresponding tube/trajectory picture obtained by translating each local mirror choice into a pairing of the four incident edges; the disjoint trajectories induce a perfect matching of the external edges.
}
\label{fig:MirrorTube}
\end{figure}

Fix $L\in\mathbb{N}$ and let
\eq
\Lambda:=\{1,2,\ldots,L\}\times\{1,2,\ldots,L\},
\en
be the $L\times L$ square lattice.
We consider the nearest-neighbor graph on $\Lambda$ with periodic boundary conditions in the second (vertical) coordinate and open boundary conditions in the first (horizontal) coordinate.
More precisely,  for $(x,y)\in\Lambda$, we connect $(x,y)$ to $(x,y\pm1)$ with $y\pm1$ understood modulo $L$, and to $(x\pm1,y)$ provided $x\pm1\in\{1,2,\ldots,L\}$.
We then attach one external edge to each boundary vertex on the left ($x=1$) and on the right ($x=L$), so that every vertex has four incident edges.
In Fig.~\ref{fig:MirrorTube}~(a) and (c), periodic boundary conditions are indicated by dotted lines, and the added external edges are depicted in thick lines.

A mirror configuration assigns to each vertex $(x,y)\in\Lambda$ one of three local scatterers:
an empty site (no mirror), a ``/'' mirror, or a ``\textbackslash'' mirror.
See Fig.~\ref{fig:MirrorTube}~(a).
A particle (or light ray) moves deterministically along edges: when it arrives at a vertex along one incident edge, it leaves along the edge prescribed by the local scatterer.

Since the scattering is time-reversal invariant, each local rule induces a pairing of the four incident edges at the vertex.
Concretely, if we label the four incident edges by the cardinal directions
$\{\mathrm{N},\mathrm{E},\mathrm{S},\mathrm{W}\}$, the three local rules are equivalent to the three pairings as follows.
See also Fig.~\ref{fig:MirrorTube}~(b).
\eq
\text{(empty)}\Longleftrightarrow\{\{\mathrm{W},\mathrm{E}\},\{\mathrm{N},\mathrm{S}\}\},\quad
\text{(/ mirror)}\Longleftrightarrow\{\{\mathrm{W},\mathrm{N}\},\{\mathrm{E},\mathrm{S}\}\},\quad
\text{(\textbackslash\ mirror)}\Longleftrightarrow\{\{\mathrm{W},\mathrm{S}\},\{\mathrm{E},\mathrm{N}\}\}.
\en

Thus, a mirror configuration is exactly the same object as an assignment, to each vertex, of one of the three pairings of its four incident edges.
See Fig.~\ref{fig:MirrorTube}~(c).
This is nothing but the definition of the model presented in the main text.

One can attempt to generalize the mirror picture to $d\ge3$ by placing geometric reflectors
(e.g., mirror planes) at each vertex and letting particles undergo specular reflection.
However, such geometric reflections realize only a restricted subset of the possible pairings
of the $2d$ incident lattice edges.  In particular, for $d=3$ there are $(2d-1)!!=15$ pairings of
the six incident edges, whereas mirror-plane reflections acting on the coordinate directions generate
only a proper subset.
For this reason, a literal ``mirror'' construction in $d\ge3$ does not coincide with the
simple and natural local-pairing model adopted in the main text, where each of the $(2d-1)!!$
pairings is chosen with equal probability.

In the main text, we also referred to numerical results for the three-dimensional Lorentz mirror model with the ``orthogonal rule,'' in which a particle is forced to turn at every visited vertex (equivalently, no edge is paired with its opposite).
We expect that the models with the standard pairing rule and the orthogonal pairing rule belong to the same universality class in $d=3$ and exhibit the same macroscopic scaling behavior.
This is in contrast with $d=2$, where the Lorentz mirror model with the
orthogonal rule is known to belong to a different universality class than the
standard model, namely that of the percolation critical point; see, e.g.,
\cite{ZiffKongCohen1991,NahumSernaSomozaOrtuno2013} and End Matter.  Thus the
percolation-critical case should be regarded as a special two-dimensional member
of the broader loop/current-matching picture, rather than as the mechanism
behind the three-dimensional results.

\section{Gaussian closure: derivation of \eqref{eq:gaussrec} and the $2/3$ law}
\label{sm:Gauss}
We record the computations behind the Gaussian heuristic quoted in the main text.

\subsection{Conditional mean/variance of the interface matching}
Fix nonnegative integers $a,b$ and consider a uniform perfect matching of $a+b$ labeled interface edges,
where $a$ edges are marked L and $b$ edges are marked R.
Let $\Lambda(a,b)$ be the number of L--R pairs created by this matching.
Introduce indicators $I_i$ ($i=1,\ldots,a$) that equal $1$ iff the $i$-th L edge is paired to an R edge.
Then $\Lambda(a,b)=\sum_{i=1}^a I_i$ and by symmetry
\eq
\sbkt{I_i}=\frac{b}{a+b-1},
\en
so
\eq
\tmu(a,b):=\sbkt{\Lambda(a,b)}=\frac{ab}{a+b-1}.
\label{e:EMell0}
\en
For $i\neq j$, after fixing one L--R matching there remain $a+b-3$ possible partners for the other L edge,
of which $b-1$ are R, hence
\eq
\sbkt{I_iI_j}=\frac{b}{a+b-1}\,\frac{b-1}{a+b-3}.
\en
Therefore $\sbkt{\Lambda(\Lambda-1)}=a(a-1)b(b-1)/\{(a+b-1)(a+b-3)\}$ and
\eq
\tv(a,b):=\Var(\Lambda(a,b))
=\frac{2ab(a-1)(b-1)}{(a+b-3)(a+b-1)^2}.
\label{e:EMtv}
\en
For $a\simeq b$ large, $\tv(a,b)=2a^2b^2/(a+b)^3\{1+O(a^{-1}+b^{-1})\}$, so in particular
$\tv(\nu,\nu)=\nu/4+O(1)$ for $\nu\gg1$.

\subsection{Input statistics at scale $n-1$}
Let $m:=2^{d-1}$.
In the recursion \eqref{eq:rec}, $\ellL=\sum_{j=1}^{m}\ell_j$ and $\ellR=\sum_{j=m+1}^{2m}\ell_j$,
where $\ell_j$ are assumed to be i.i.d.\ with Gaussian law $P_{n-1}$ with parameters $(\mu_{n-1},v_{n-1})$.
Thus
\eq
\sbkt{\ellL}=\sbkt{\ellR}=m\mu_{n-1},\qquad
\Var(\ellL)=\Var(\ellR)=m v_{n-1},
\en
and $\ellL$ and $\ellR$ are independent.

\subsection{Mean recursion for $d\ge3$}
Under the Gaussian closure, we assume that $P_{n-1}$ is a (discrete) Gaussian with mean $\mu_{n-1}$ and variance $v_{n-1}$ of the same order of magnitude.  We find an approximate recursion relation for  $\mu_n=\sbkt{\tmu(\ellL,\ellR)}$ by expanding
$\tmu(a,b)=ab/(a+b-1)$ around the symmetric point $(a,b)=(m\mu_{n-1},m\mu_{n-1})$.
We assume $\mu_{n-1}$ large and the leading term gives
\begin{eqnarray}
\mu_n &\simeq & \tmu(m\mu_{n-1},m\mu_{n-1})\nonumber\\
&=&\frac{m^2\mu_{n-1}^2}{2m\mu_{n-1}-1}\simeq \frac{m}{2}\mu_{n-1}
=2^{d-2}\mu_{n-1}.
\label{e:EMmurec}
\end{eqnarray}

\subsection{Variance recursion for $d\ge3$}
By the law of total variance,
\eq
v_n=\sbkt{\tv(\ellL,\ellR)}+\Var(\tmu(\ellL,\ellR)).
\en
For the intrinsic term, evaluating \eqref{e:EMtv} at $(m\mu_{n-1},m\mu_{n-1}) $ yields
\eq
\sbkt{\tv(\ellL,\ellR)}\simeq \tv(m\mu_{n-1},m\mu_{n-1})
\simeq \frac{m\mu_{n-1}}{4}=2^{d-3}\mu_{n-1}.
\label{e:EMv_intr}
\en
For the propagated term, linearize $\tmu(a,b)$ at symmetry:
\eq
\frac{\partial\tmu(a,b)}{\partial a}=\frac{b(b-1)}{(a+b-1)^2},\qquad
\frac{\partial\tmu(a,b)}{\partial b}=\frac{a(a-1)}{(a+b-1)^2},
\en
so at $a=b=m\mu_{n-1}$ one has $\partial\tmu/\partial a=\partial\tmu/\partial b=1/4+O(\mu_{n-1}^{-1})$.
Using independence of $\ellL,\ellR$,
\begin{eqnarray}
\Var(\tmu(\ellL,\ellR))
&\simeq & \Bigl(\frac14\Bigr)^2\{\Var(\ellL)+\Var(\ellR)\}\nonumber\\
&=&\frac{1}{16}(mv_{n-1}+mv_{n-1})\nonumber\\
&=&\frac{m}{8}v_{n-1}
=2^{d-4}v_{n-1}.
\label{e:EMv_prop}	
\end{eqnarray}

Combining \eqref{e:EMv_intr} and \eqref{e:EMv_prop} yields \eqref{eq:gaussrec}.
Dividing by \eqref{e:EMmurec} yields \eqref{eq:ratio23}, hence $v_n/\mu_n\to2/3$.

\subsection{The marginal case $d=2$}
Here $m=2$ and the leading term in \eqref{e:EMmurec} only gives $\mu_n\simeq\mu_{n-1}$, so we keep the
next-order corrections.
First,
\eq
\tmu(2\mu,2\mu)=\frac{4\mu^2}{4\mu-1}=\mu+\frac14+O(\mu^{-1}).
\en
Second, a Taylor expansion of $\tmu(a,b)=ab/(a+b-1)$ around $(2\mu,2\mu)$ shows that averaging over the
fluctuations of $\ellL,\ellR$ produces a correction of order $v/\mu$, and one finds
\eq
\mu_n-\mu_{n-1}\simeq \frac{1}{4}-\frac{v_{n-1}}{4\mu_{n-1}}.
\label{e:EMmu2}
\en
Meanwhile, the variance recursion remains
\eq
v_n\simeq \frac{\mu_{n-1}}{2}+\frac{v_{n-1}}{4},
\label{e:EMv2}
\en
so $v_n/\mu_n\to2/3$ as before. Substituting $v_{n-1}/\mu_{n-1}\simeq2/3$ into \eqref{e:EMmu2} yields
$\mu_n-\mu_{n-1}\simeq 1/12$, hence $\mu_n\simeq n/12=(\log L_n)/(12\log2)$ and
$v_n\simeq (2/3)\mu_n\simeq n/18$.

\subsection{Universal laws for higher cumulants}
One may go beyond the Gaussian closure by allowing cubic and quartic corrections in the
effective potential $\log P_n(\ell)$.
A convenient way is to expand in the centered variable
$\varphi:=(\ell-\mu_n)/\sqrt{v_n}$, i.e., to supplement the Gaussian form with $\varphi^3$
and $\varphi^4$ terms and carry out a standard perturbative analysis of the recursion.
This predicts that, for any $d\ge2$, the third and fourth \emph{cumulants} of the crossing number
$\hell$ obey the universal asymptotics (see [SM1])
\eq
\kappa^{(3)}_n:=\sbkt{(\hell-\mu_n)^3}_n \simeq \frac{4}{15}\,\mu_n,
\qquad
\kappa^{(4)}_n:=\sbkt{(\hell-\mu_n)^4}_n-3v_n^2 \simeq -\frac{8}{105}\,\mu_n,
\en
as $n\uparrow\infty$.
Equivalently, the ratios $\kappa^{(3)}_n/\mu_n$ and $\kappa^{(4)}_n/\mu_n$ approach the
universal constants $4/15$ and $-8/105$, respectively, in the same sense as the $2/3$ law for
$v_n/\mu_n$.

These higher-cumulant laws provide a stringent numerical test of universality for the
(non-hierarchical) original Lorentz mirror model, especially in $d=3$.
For $d=2$, the ``$4/15$ law'' was already obtained from a field-theoretic RG analysis and
confirmed numerically in \cite{NahumSernaSomozaOrtuno2013}.

\section{Extension of the proof of Theorem~\ref{lowergen} to general $d\ge3$}
\label{sm:Theorems3}
This section supplies the details omitted in End Matter and derives the $d$-dependent constant $C_d$ in Theorem~\ref{lowergen}.
We keep the notation of End Matter.
Set $m=2^{d-1}$ as above.  Note that $m=4$ in $d=3$.
Recall the definitions of $\eta_n$ and $u_n$ in \eqref{eq:etadef} and \eqref{eq:un}, and the identity \eqref{eq:eta=u}.

\subsection{A concavity bound on $u_n$}
From \eqref{eq:un} and \eqref{eq:exp_n}, since $(\hellL+1)^{-1}$ depends only on the $m=2^{d-1}$ variables in the left half, we may write
\eq
u_n
=\sum_{\ell_1,\ldots,\ell_m}\Bigl(1+\sum_{j=1}^{m}\ell_j\Bigr)^{-1}\prod_{j=1}^{m}P_n(\ell_j),
\en
where the sum is over nonnegative even $\ell_1,\ldots,\ell_m$.
Make the change of variables $y=(\ell+1)^{-1}\in(0,1]$ and define
$Q_n(y):=P_n(y^{-1}-1)$.
Then $\eta_n=\sum_y y\,Q_n(y)$ and the above becomes
\eq
u_n
=\sum_{y_1,\ldots,y_m}\Bigl(\sum_{j=1}^m y_j^{-1}-(m-1)\Bigr)^{-1}\prod_{j=1}^{m}Q_n(y_j).
\label{eq:un_y_general}
\en
Fix $y_2,\ldots,y_m\in(0,1]$ and set
$c:=\sum_{j=2}^m y_j^{-1}-(m-1)\ge0$.
Then
\eq
(y_1^{-1}+c)^{-1}=\frac{y_1}{1+c y_1}
\en
is concave in $y_1\in(0,1]$, since its second derivative equals $-2c(1+c y_1)^{-3}\le0$.
Applying Jensen's inequality to the $y_1$-average in \eqref{eq:un_y_general} yields
\eq
u_n\le
\sum_{y_2,\ldots,y_m}\Bigl(\eta_n^{-1}+\sum_{j=2}^m y_j^{-1}-(m-1)\Bigr)^{-1}\prod_{j=2}^{m}Q_n(y_j).
\en
Repeating the same argument for $y_2,\ldots,y_m$ gives
\eq
u_n\le \bigl(m\eta_n^{-1}-(m-1)\bigr)^{-1}
=\frac{\eta_n}{m-(m-1)\eta_n}
=\frac{\eta_n}{2^{d-1}-(2^{d-1}-1)\eta_n}.
\label{eq:u_eta_general}
\en

\subsection{A quadratic recursion for $\eta_n$}
Substituting \eqref{eq:u_eta_general} (with $n\mapsto n-1$) into \eqref{eq:eta=u} and using that
$x\mapsto 2x-x^2$ is increasing on $[0,1]$, we obtain
\eq
\eta_n
\le
\frac{2m\,\eta_{n-1}-(2m-1)\eta_{n-1}^2}{\bigl(m-(m-1)\eta_{n-1}\bigr)^2}.
\label{eq:eta_rational_general}
\en
A direct computation shows that for any $\eta\in[0,1]$,
\eqa
&\frac{2m\,\eta-(2m-1)\eta^2}{\bigl(m-(m-1)\eta\bigr)^2}
-\Bigl(\frac{2}{m}\eta+\Bigl(\frac{m-1}{m}\Bigr)^2\eta^2\Bigr)
\nl
&\qquad=
-\frac{\eta^2\bigl(\eta(m-1)^2-m(m-2)\bigr)^2}{m^2\bigl(m-(m-1)\eta\bigr)^2}
\le0.
\label{eq:eta_alg_general}
\ena
Combining \eqref{eq:eta_rational_general} and \eqref{eq:eta_alg_general} yields the quadratic recursion
\eq
\eta_n\le \frac{2}{m}\eta_{n-1}+\Bigl(\frac{m-1}{m}\Bigr)^2\eta_{n-1}^2
=2^{2-d}\eta_{n-1}+(1-2^{1-d})^2\eta_{n-1}^2.
\label{eq:eta_rec_general}
\en

\subsection{Iteration and the constant $C_d$}
Let $\alpha:=2/m=2^{2-d}$ and set
\eq
x_n:=\alpha^{-n}\eta_n=\Bigl(\frac{m}{2}\Bigr)^n\eta_n=2^{(d-2)n}\eta_n.
\en
Then \eqref{eq:eta_rec_general} implies
\eq
x_n\le x_{n-1}+\beta\,\alpha^{\,n-2}x_{n-1}^2,
\qquad
\beta:=\Bigl(\frac{m-1}{m}\Bigr)^2,
\en
and hence
\eq
\frac{1}{x_n}\ge \frac{1}{x_{n-1}}-\beta\,\alpha^{\,n-2},
\en
since $(x+a x^2)^{-1}\ge x^{-1}-a$ for $x>0$ and $a\ge0$.
Summing and using $x_0=\eta_0=(A_0+1)^{-1}$ yields
\eq
\frac{1}{x_n}
\ge
(A_0+1)-\beta\sum_{j=1}^{n}\alpha^{j-2}
\ge
(A_0+1)-\frac{\beta}{\alpha(1-\alpha)}
=
(A_0+1)-\frac{(m-1)^2}{2(m-2)}.
\label{eq:x_lower_general}
\en
Therefore
\eq
\eta_n^{-1}
=\alpha^{-n}x_n^{-1}
\ge
\Bigl(A_0+1-\frac{(m-1)^2}{2(m-2)}\Bigr)\Bigl(\frac{m}{2}\Bigr)^n
=
\Bigl(A_0+1-\frac{(2^{d-1}-1)^2}{2(2^{d-1}-2)}\Bigr)2^{(d-2)n}.
\en
Since $A_n/L_n=A_0\,2^{(d-2)n}$ and $\mu_n\ge \eta_n^{-1}-1$, we conclude
\eq
\mu_n\ge
\Biggl[
1-\frac{1}{A_0}\biggl(\frac{(2^{d-1}-1)^2}{2(2^{d-1}-2)}-1\biggr)
\Biggr]\frac{A_n}{L_n}-1,
\en
which is exactly the bound stated in Theorem~\ref{lowergen} (with the constant $C_d$ given there).

\section{Numerical simulations}
\label{SM:NumericsHierarchical}
The present section provides additional information on our numerical work. 
The codes are archived on Zenodo (DOI: 10.5281/zenodo.18622214).
We focus on the hierarchical Lorentz mirror model, where the recursion \eqref{eq:rec} can be iterated essentially exactly (i.e., without Monte Carlo sampling), and we briefly discuss the original (non-hierarchical) Lorentz mirror model in the final subsection.

\subsection{The set-up and codes for the hierarchical recursion}
Recall that $P_n(\ell)$ (with $\ell\in2\mathbb{Z}_{\ge0}$) denotes the probability that the generation-$n$ block has exactly $\ell$ crossings.
It is determined by the recursion \eqref{eq:rec}, where the kernel $K(\ell\,|\,\ellL,\ellR)$ is given by \eqref{eq:kernel}.

We provide two codes for iterating the recursion:
\begin{itemize}
\item \texttt{scramblers.cpp} (2D): iterates \eqref{eq:rec} with $d=2$, i.e.\ $m:=2^{d-1}=2$ subblocks in parallel per half.
\item \texttt{scramblers3d.cpp} (3D): iterates \eqref{eq:rec} with $d=3$, i.e.\ $m:=2^{d-1}=4$ subblocks in parallel per half.
\end{itemize}

To be more precise, we first compute the $m$-fold convolution
\eq
P_{n-1}^{*m}(\ell'):=\mathop{\sum_{\ell_1,\ldots,\ell_m}}_{(\sum_{j=1}^m\ell_j=\ell')}\prod_{j=1}^mP_{n-1}(\ell_j),
\en
and then evaluate the next-generation distribution via
\eq
P_n(\ell)=\sum_{\ellL,\ellR}K(\ell\,|\,\ellL,\ellR)\,P_{n-1}^{*m}(\ellL)\,P_{n-1}^{*m}(\ellR).
\en
In both codes, the full law on even integers is stored as an array,
$\texttt{prob[k]}=\operatorname{Prob}(\hell_n=2k)$, and the recursion \eqref{eq:rec} is evaluated directly.
Very small probabilities (below a pruning threshold $\simeq10^{-18}$) are discarded to keep the support finite.
In our implementation, the total discarded mass is negligible at the scale of the reported observables; after pruning we renormalize the distribution to have unit total mass.

The codes output the mean $\mu_n=\sbkt{\hell}_n$, the variance $v_n=\sbkt{(\hell-\mu_n)^2}_n$, and their ratio $v_n/\mu_n$.
No Monte Carlo sampling is used: apart from the pruning step just described, the computation is exact.

As in the main text, we set the initial distribution as $P_0(\ell)=\delta_{\ell,A_0}$, where $A_0$ is a positive even integer.
In the $d=3$ plots below we use $A_0=2$, while in the $d=2$ plots we use $A_0=90$ (see the discussion below).

\begin{figure}[t]
\centering
\includegraphics[width=0.49\linewidth]{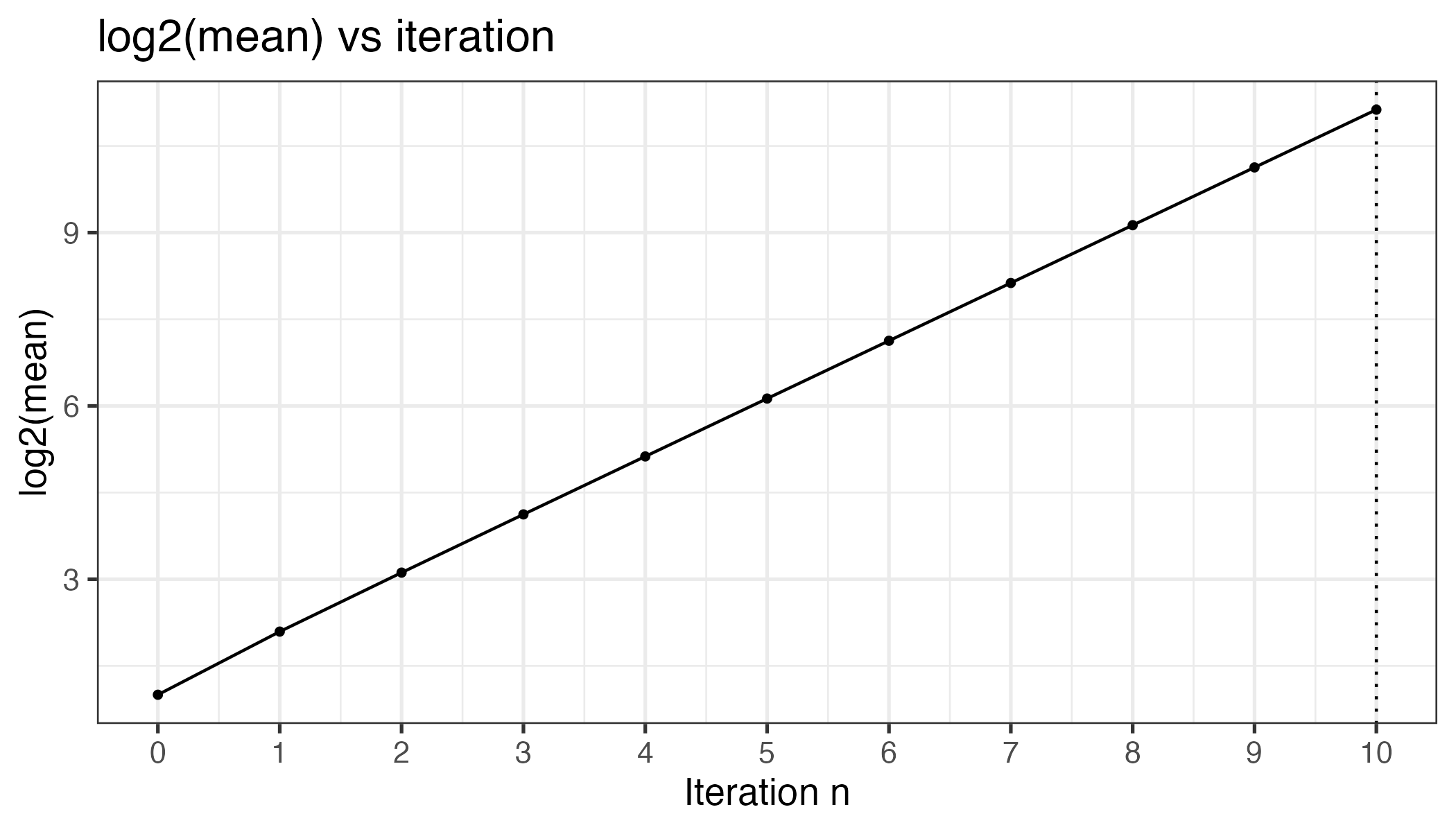}
\includegraphics[width=0.49\linewidth]{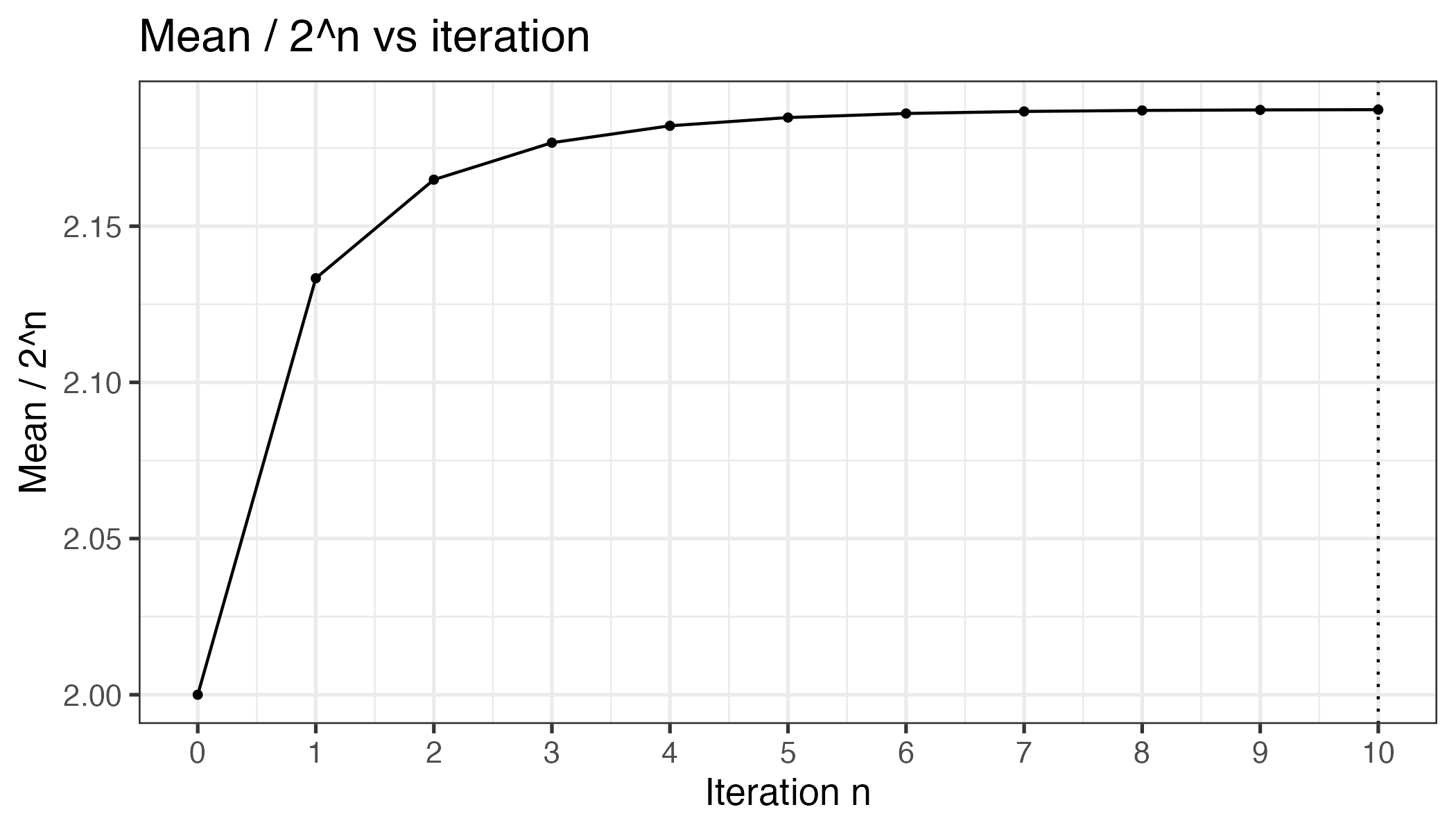}
\caption{(3D Hierarchical model) Mean conductance $\mu_n$ in $d=3$ with initial crossing $A_0=2$.
Left: $\log_2\mu_n$ versus iteration $n=\log_2 L_n$.
The linear growth is consistent with the normal-transport scaling $\mu_n\propto L_n=2^n$.
Right: As $n$ increases, the ratio $\mu_n/2^n$ converges to a constant (the conductivity).}
\label{fig:3D-H-mu}
\end{figure}

\begin{figure}[t]
\centering
\includegraphics[width=0.5\linewidth]{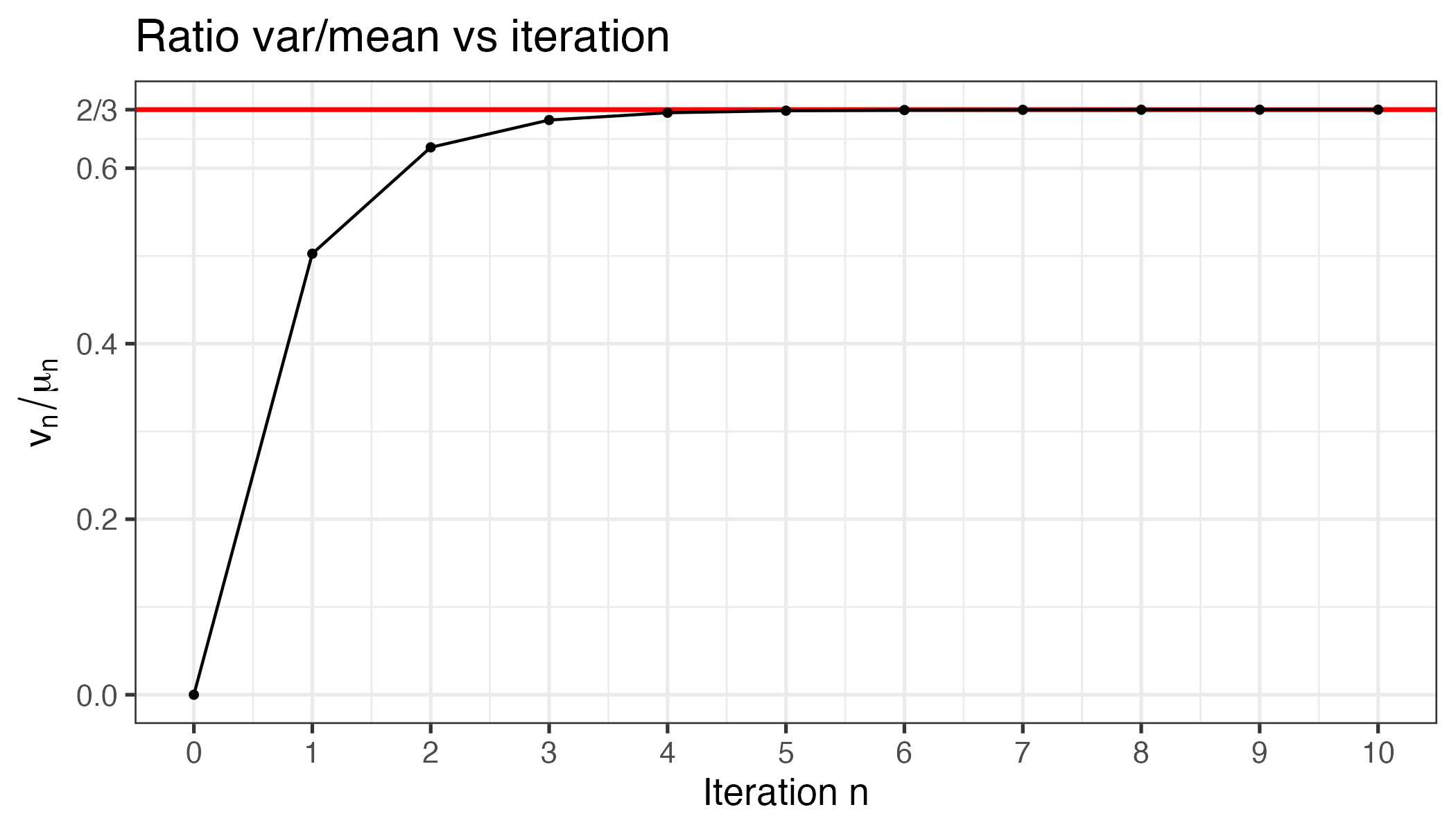}
\caption{(3D Hierarchical model) Ratio $v_n/\mu_n$ versus iteration $n=\log_2 L_n$ in $d=3$ with initial crossing $A_0=2$.
The ratio converges quickly toward $2/3$ (thick red line).}
\label{fig:SM-3D-ratio}
\end{figure}

\subsection{Results in $d=3$: normal-conductance scaling and the $2/3$ law}
In $d=3$, we proved that the mean conductance $\mu_n$ grows proportionally to $2^n$ for any $A_0\ge2$.
Here we choose the smallest initial crossing number and set $A_0=2$.

Figure~\ref{fig:3D-H-mu} shows $\log_2\mu_n$ and $\mu_n/2^n$ as functions of the generation $n$.
The behavior is consistent with normal-transport scaling.
Note that the conductivity $\kappa:=\lim_{n\uparrow\infty}\mu_n/L_n$ (read off from the right panel) is nonuniversal and depends on the initial distribution $P_0$.

From the Gaussian closure, we expect the variance-to-mean ratio $v_n/\mu_n$ to converge to the universal constant $2/3$.
Figure~\ref{fig:SM-3D-ratio} shows $v_n/\mu_n$ for $n\le10$.
The ratio rapidly approaches a plateau very close to $2/3$ (already by $n\simeq6$), providing clean numerical support for the $2/3$ mean--variance law in $d=3$.

\begin{figure}[t]
\centering
\includegraphics[width=0.5\linewidth]{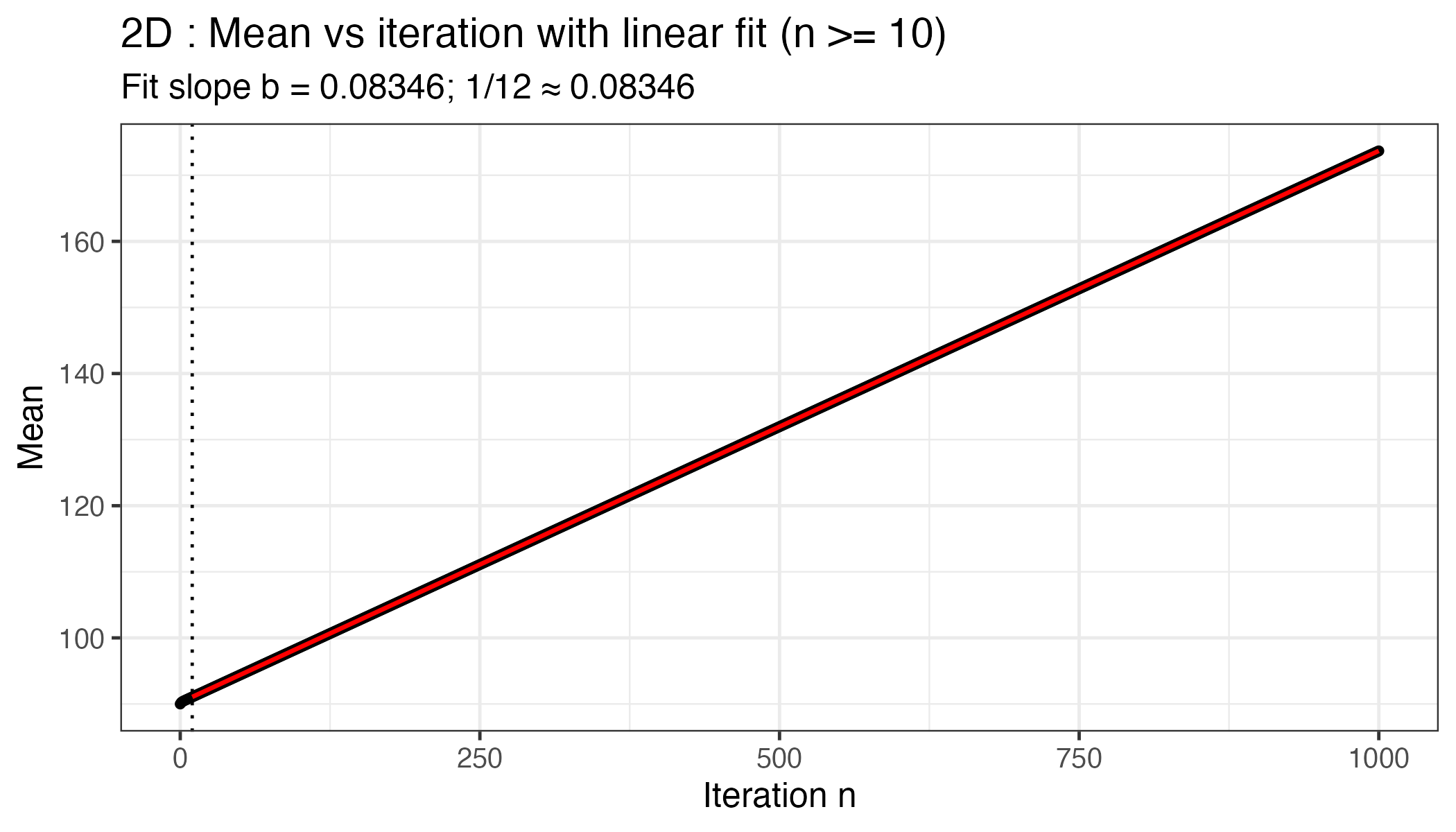}
\caption{(2D Hierarchical model) Mean conductance $\mu_n=\sbkt{\hell}_n$ versus iteration $n=\log_2 L_n$ in $d=2$ with initial crossing $A_0=90$.
A linear regression fit for $n\ge10$ yields the slope $b\simeq0.08346$, which is indistinguishable from $1/12$.}
\label{fig:SM-2D-mean}
\end{figure}

\begin{figure}[t]
\centering
\includegraphics[width=0.5\linewidth]{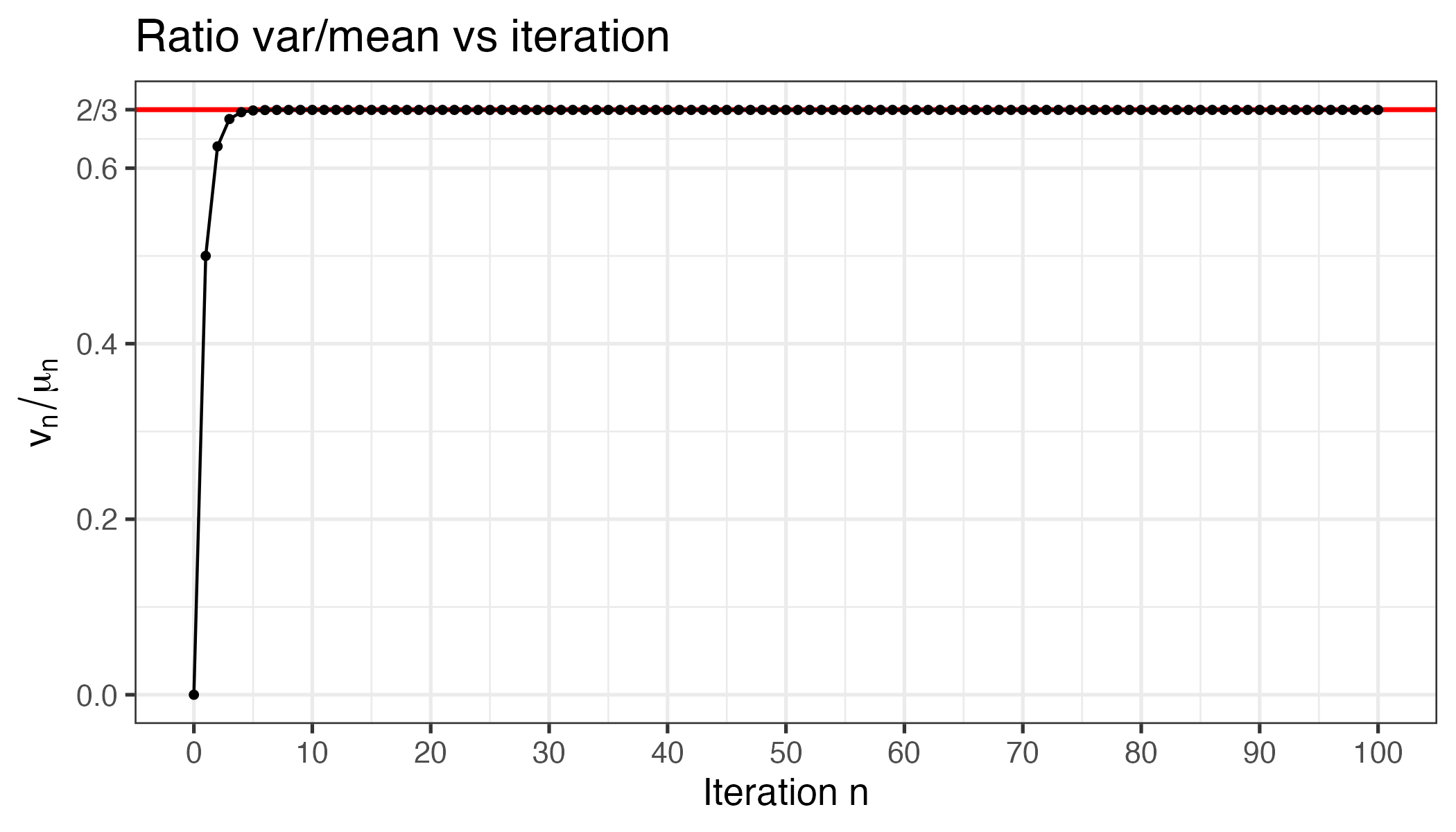}
\caption{(2D Hierarchical model) Ratio $v_n/\mu_n$ versus iteration $n=\log_2 L_n$ in $d=2$ with initial crossing $A_0=90$.
The thick red line marks the target value $2/3$.}
\label{fig:SM-2D-ratio}
\end{figure}

\subsection{Results in $d=2$: logarithmic growth and the $2/3$ law}
In the marginal dimension $d=2$, the Gaussian closure predicts a slow growth of the mean conductance, $\mu_n\simeq n/12$, equivalently $\mu_n\simeq(\log L_n)/(12\log 2)$, as in \eqref{eq:2D}.

If we start from the smallest crossing number $A_0=2$, we observe the expected slow growth, but the distribution $P_n(\ell)$ is then essentially supported only on small values of $\ell$.
(For example, $P_{100}(\ell)$ is almost negligible for $\ell\ge20$.)
In this regime, measures of ``distance to Gaussianity'' are strongly affected by discreteness.

To reduce such discreteness effects, we choose here a larger initial crossing number, $A_0=90$.
Then, after a single application of the recursion \eqref{eq:rec}, we obtain
$P_1(\ell)=K(\ell\,|\,180,180)$, which is already close to a Gaussian by the asymptotics of the kernel.

Figure~\ref{fig:SM-2D-mean} shows $\mu_n$ versus $n$, together with a linear regression for $n\ge10$.
The best-fit slope is
\eq
\mu_n \simeq a+b\,n\quad \text{(as $n\uparrow\infty$)}, \qquad b \simeq 0.08346,
\en
which essentially coincides with the Gaussian-closure prediction $b=1/12=0.08333\cdots$.

The variance-to-mean ratio $v_n/\mu_n$ is also well behaved.
As shown in Fig.~\ref{fig:SM-2D-ratio}, $v_n/\mu_n$ stabilizes near the universal value $2/3$ after only a short transient.

\subsection{Convergence to Gaussianity in $d=3$}
\label{SM:Gaussianity}
\begin{figure}[t]
\centering
\includegraphics[width=0.49\linewidth]{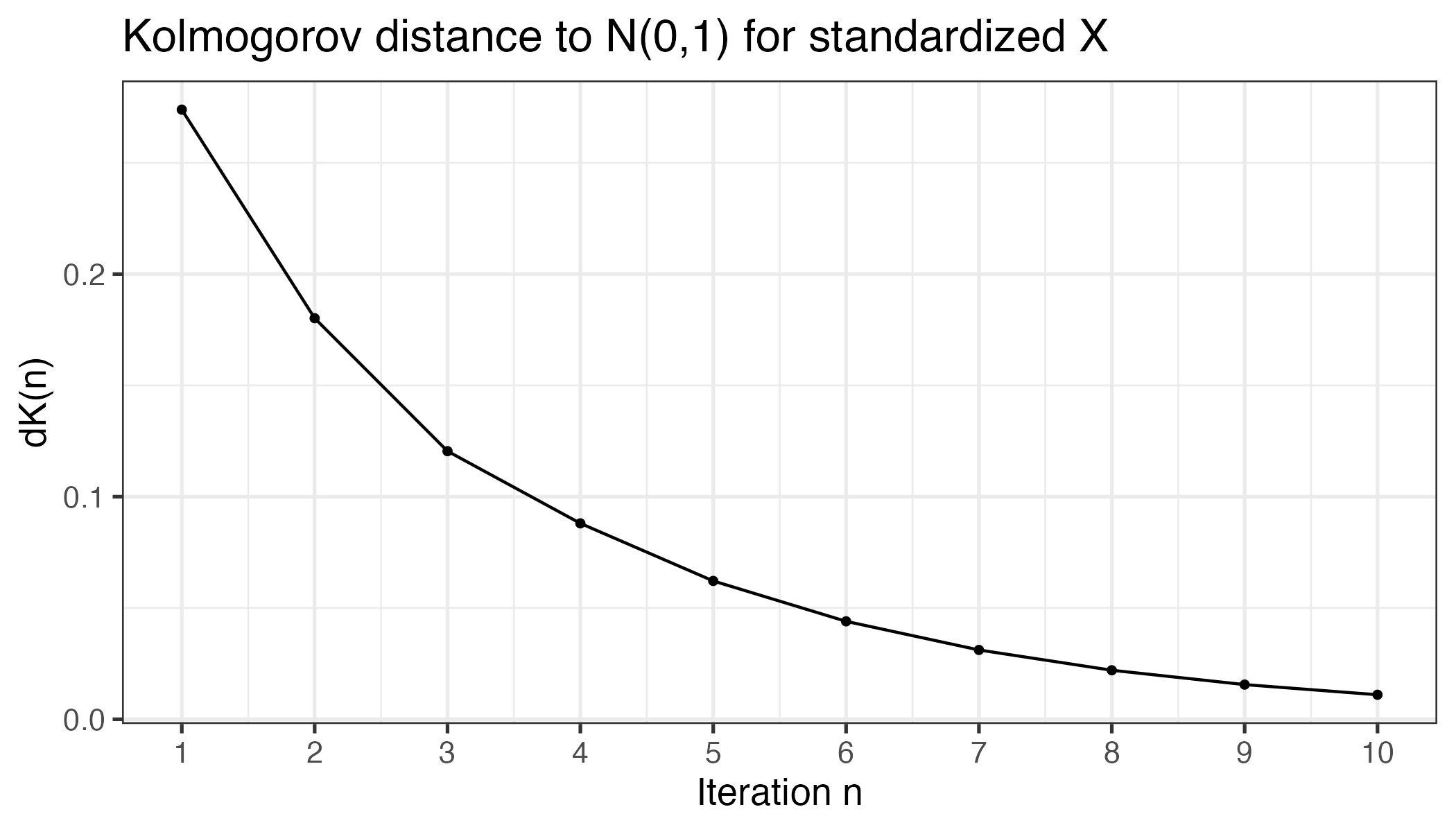}
\includegraphics[width=0.49\linewidth]{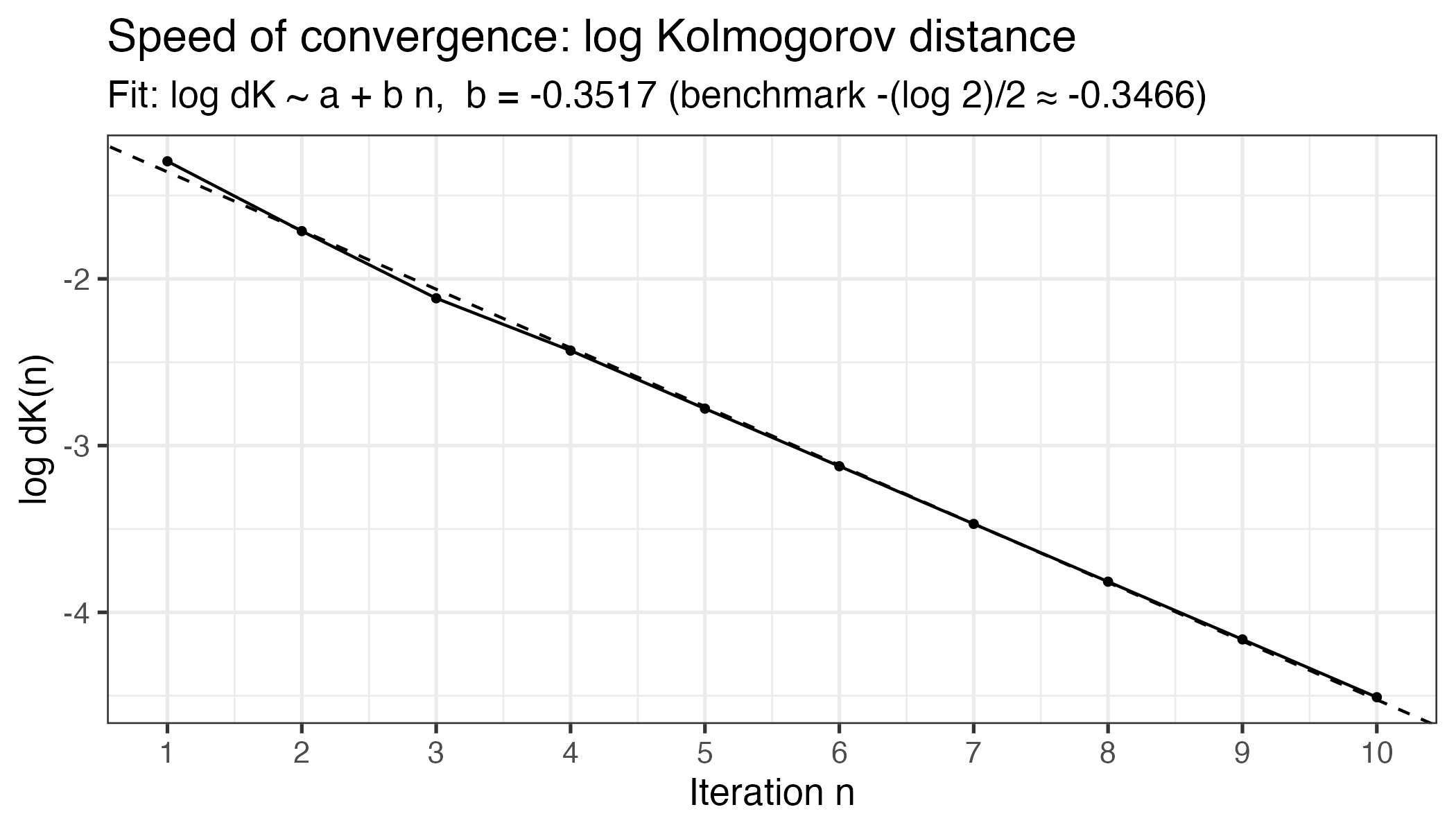}
\caption{(3D Hierarchical model) Left: Kolmogorov distance $\dK(n)$ defined in \eqref{SM:Kolmogorov} between the standardized iterate $Z_n$ and $\mathcal N(0,1)$.
The distance decreases rapidly with $n$.
Right: $\log \dK(n)$ versus $n$ with a linear fit.
The fitted slope (dashed line) $b\simeq -0.3517$ is close to $-(\log 2)/2\simeq -0.3466$, suggesting $\dK(n)\simeq 2^{-n/2}$ as $n\uparrow\infty$.}
\label{fig:SM-3D-K}
\end{figure}
As mentioned in the main text, for the hierarchical model with $d=3$ the distribution $P_n$ rapidly approaches a Gaussian shape as $n$ grows.
To quantify this, we use the Kolmogorov distance between the standardized crossing number and the standard normal distribution.

Let $\hell_n$ denote the number of crossings in the generation-$n$ block and define the standardized variable
\eq
Z_n:=\frac{\hell_n-\mu_n}{\sqrt{v_n}}.
\en
Let $F_n(z)=\operatorname{Prob}[Z_n\le z]$ be the cumulative distribution function (CDF) of $Z_n$, and let $\Phi(z)$ be the CDF of the standard normal $\mathcal N(0,1)$.
We define the Kolmogorov distance by
\eq
\dK(n):=\sup_{z\in\mathbb{R}}\,\bigl|F_n(z)-\Phi(z)\bigr|.
\label{SM:Kolmogorov}
\en

As shown in the left panel of Fig.~\ref{fig:SM-3D-K}, $\dK(n)$ decreases quickly over the first few iterations and reaches the percent level by $n\simeq10$.
A linear fit of $\log \dK(n)$ versus $n$ (right panel) yields
\eq
\log \dK(n)\simeq a + b\,n\quad\text{(as $n\uparrow\infty$)},\qquad b\simeq -0.3517.
\en
This demonstrates exponential convergence of the standardized law to $\mathcal N(0,1)$ in the $d=3$ hierarchical model.

Noting that the estimated slope is close to $-(\log 2)/2$, we are led to conjecture
\eq
\dK(n)\simeq C\,2^{-n/2}\propto\frac{1}{\sqrt{\mu_n}},
\en
as $n\uparrow\infty$.

\subsection{Slow convergence to Gaussianity in $d=2$}
\label{SM:Gaussianity2}
\begin{figure}[t]
\centering
\includegraphics[width=0.49\linewidth]{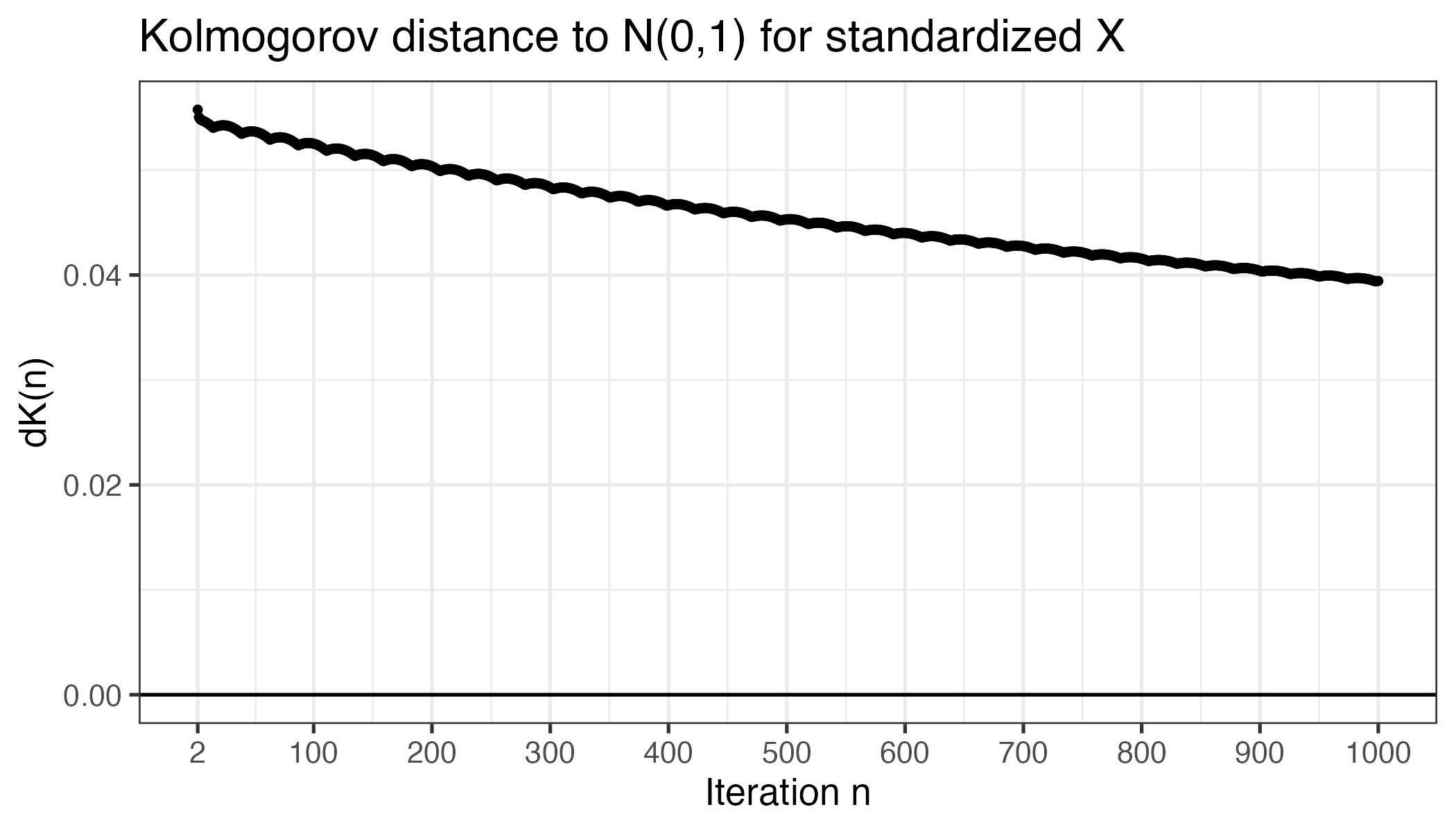}
\includegraphics[width=0.49\linewidth]{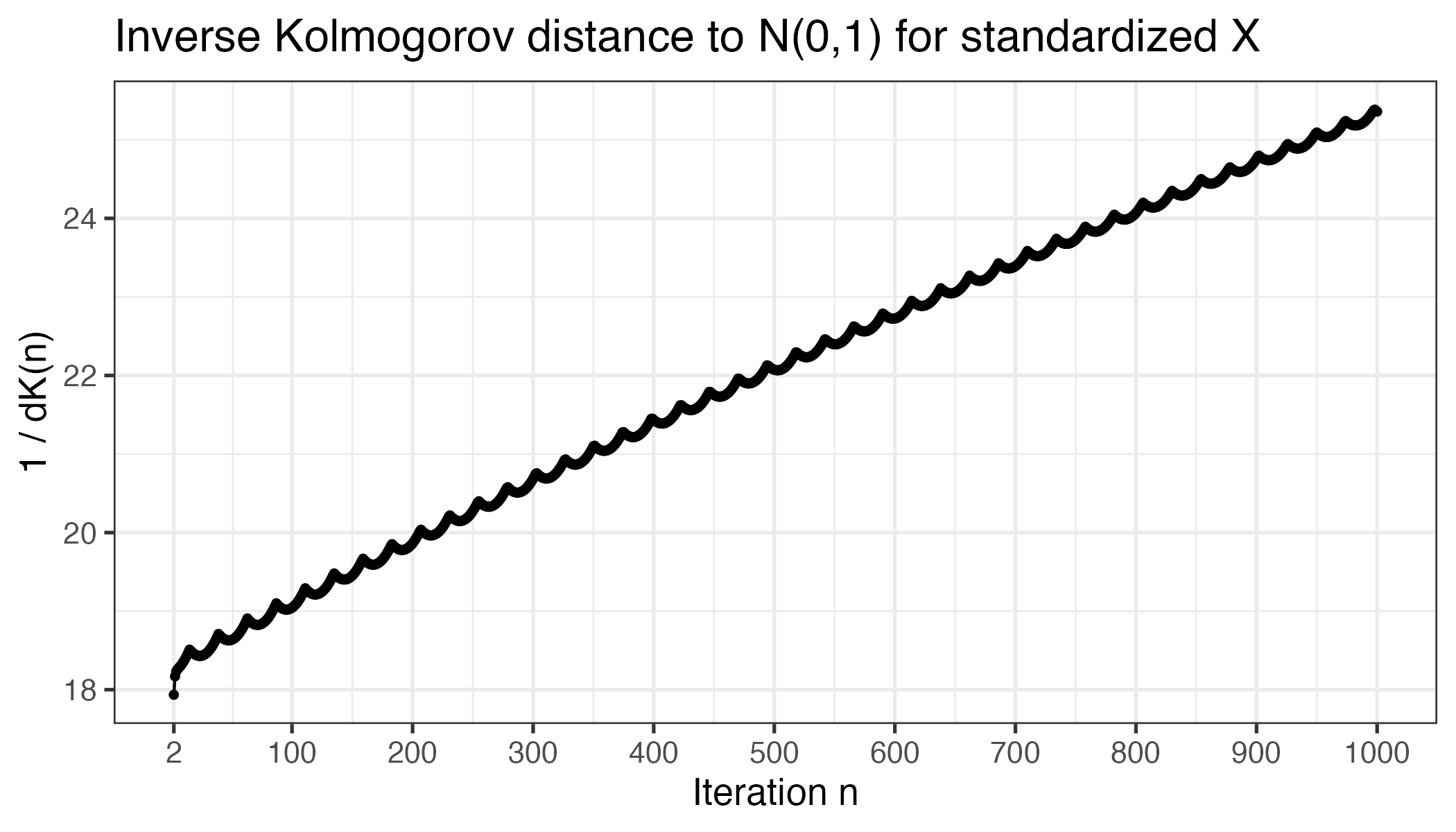}
\caption{(2D Hierarchical model) Left: Kolmogorov distance $\dK(n)$ defined in \eqref{SM:Kolmogorov} between the standardized iterate $Z_n$ and $\mathcal N(0,1)$.
The distance decreases slowly with $n$.
Right: $1/\dK(n)$ versus $n$ exhibits an approximately linear growth, indicating $\dK(n)\propto n^{-1}$ at large $n$.
Mild oscillations visible in the data are due to the lattice nature of $\ell$ and are not essential.}
\label{fig:SM-2D-K}
\end{figure}

Let us turn to the case $d=2$, where we set $A_0=90$.
Recall that $P_1(\ell)=K(\ell\,|\,180,180)$ is already close to a Gaussian.

As can be seen in the left panel of Fig.~\ref{fig:SM-2D-K}, the Kolmogorov distance $\dK(n)$ decays very slowly with $n$.
The plot of $1/\dK(n)$ in the right panel suggests that $\dK(n)$ decays proportionally to $1/n$ for large $n$.

We shall study the approach of $P_n(\ell)$ to a Gaussian perturbatively in [SM1].

\subsection{The original Lorentz mirror model in $d=3$}
\label{sm:num3D}

\begin{figure}[t]
\centering
\includegraphics[width=0.5\linewidth]{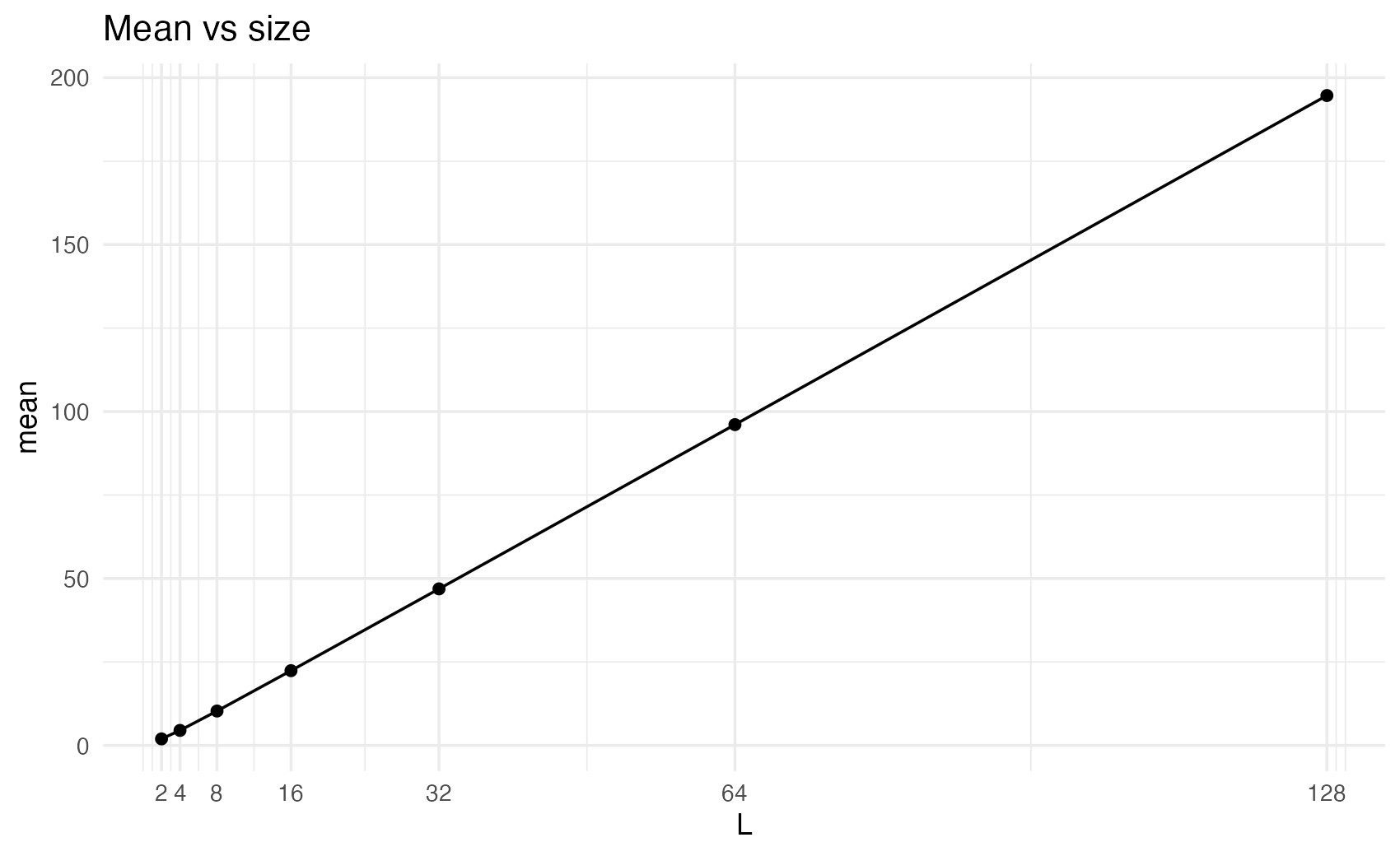}
\caption{(3D Original model) Mean conductance $\bar{\mu}(L)$ versus system size $L$ in the original Lorentz mirror model in $d=3$.
Our preliminary laptop-scale simulations are consistent with the linear growth predicted by normal transport \eqref{eq:Fick}.
For comparison, Fig.~6 of \cite{ChiffaudelLefevere} reports large-scale simulations of $\bar{\mu}(L)/L$ for $L$ from $5$ to $420$.}
\label{fig:Orgmu}
\end{figure}

As explained in the main text, we performed preliminary simulations of the original Lorentz mirror model on the $L\times L\times L$ cubic lattice for $L=2^n$ with $n=1,\ldots,7$.
Our codes (which generate quenched random environments with local pairings and count the resulting number of crossings) are:
\begin{itemize}
\item \verb!cross_stats.cpp! and \verb!cross_stats_rect.cpp!.
\end{itemize}
They are archived on Zenodo (DOI: 10.5281/zenodo.18622214).
For each $L$ and each of the two local pairing rules described in the main text, we generated $M=6.4\times 10^5$ independent environments.
For a given environment, we faithfully computed the conductance $\calC$ (the number of left--right crossings) by tracing deterministically the trajectories of particles injected from all $L^2$ external edges on the left boundary and recording their exit edges.
From these data we estimated the mean $\bar{\mu}(L)=\sbkt{\calC}$ and the variance $\bar{v}(L)=\sbkt{(\calC-\bar{\mu}(L))^2}$, together with $95\%$ confidence intervals.

Uncertainties are obtained by a block (batching) analysis: the $M$ environments are split into $K$ blocks, and the dispersion of block-level estimators provides the standard error and corresponding normal confidence intervals.
Since the generation of an environment requires specifying $L^3$ local pairings and the subsequent trajectory tracing is of comparable cost, the runtime increases rapidly with $L$; for $L=128$ the total computation time is about $2.5\times 10^4\,\mathrm{s}$ on a laptop.

The key outcome of the simulations is the approach of $\bar{v}(L)/\bar{\mu}(L)$ to the universal value $2/3$, shown in Fig.~\ref{fig:ratio3D_mirrors_panel} of the main text.
Figure~\ref{fig:Orgmu} provides complementary evidence that our preliminary simulations are consistent with the normal-transport scaling \eqref{eq:Fick} in $d=3$.
We stress that decisive numerical evidence for \eqref{eq:Fick} is provided by the large-scale simulations in \cite{ChiffaudelLefevere}.

To test the robustness of the $2/3$ law against changes in the aspect ratio,
we also performed numerical simulations with the standard pairing rule on an
$L\times 2L\times 2L$ cuboid, with transport along the short direction of
length $L$.
For $L=64$, we obtain
$\bar v(L)/\bar\mu(L)\simeq 0.663$,
in excellent agreement with the conjectured value $2/3$.

\medskip
\noindent
{\bf Reference}
[SM1] R. Lefevere and H. Tasaki, ``Hierarchical Lorentz mirror model: a perturbative approach'', in preparation.
 
\end{document}